\documentclass[twocolumn]{aastex6}
\usepackage{natbib}
\usepackage{amsmath}
\usepackage{perpage} 
\MakePerPage{footnote}

\begin{document}
\title{Complex organic molecules towards embedded low-mass protostars}

\author{Jennifer B. Bergner\altaffilmark{1}, Karin I. \"Oberg\altaffilmark{2}, Robin T. Garrod\altaffilmark{3}, Dawn M. Graninger\altaffilmark{2}}

\altaffiltext{1}{Harvard University Department of Chemistry and Chemical Biology, Cambridge, MA 02138, USA}
\altaffiltext{2}{Harvard-Smithsonian Center for Astrophysics, Cambridge, MA 02138, USA}
\altaffiltext{3}{University of Virginia Departments of Chemistry and Astronomy, Charlottesville, VA 22904, USA}

\begin{abstract}
Complex organic molecules (COMs) have been observed towards several low-mass young stellar objects (LYSOs).  Small and heterogeneous samples have so far precluded conclusions on typical COM abundances, as well as the origin(s) of abundance variations between sources.  We present observations towards 16 deeply embedded (Class 0/I) low-mass protostars using the IRAM 30m telescope.  We detect CH$_2$CO, CH$_3$CHO, CH$_3$OCH$_3$, CH$_3$OCHO, CH$_3$CN, HNCO, and HC$_3$N towards 67\%, 37\%, 13\%, 13\%, 44\%, 81\%, and 75\% of sources respectively.  Median column densities derived using survival analysis range between 6.0x10$^{10}$ cm$^{-2}$ (CH$_3$CN) and 2.4x10$^{12}$ cm$^{-2}$ (CH$_3$OCH$_3$) and median abundances range between 0.48\% (CH$_3$CN) and 16\% (HNCO) with respect to CH$_3$OH. Column densities for each molecule vary by about one order of magnitude across the sample.  Abundances with respect to CH$_3$OH are more narrowly distributed, especially for oxygen-bearing species.  We compare observed median abundances with a chemical model for low-mass protostars and find fair agreement, although some modeling work remains to bring abundances higher with respect to CH$_3$OH.  Median abundances with respect to CH$_3$OH in LYSOs are also found to be generally comparable to observed abundances in hot cores, hot corinos, and massive young stellar objects.  Compared with comets, our sample is comparable for all molecules except HC$_3$N and CH$_2$CO, which likely become depleted at later evolutionary stages.  
\end{abstract}


\section{Introduction}
{\let\thefootnote\relax\footnote{Based on observations carried out under project numbers 003-14 and 006-13 with the IRAM 30m Telescope. IRAM is supported by INSU/CNRS (France), MPG (Germany) and IGN (Spain)}}
Complex organic molecules (COMs), hydrogen-rich molecules with 6 or more atoms, have been observed towards high- and low-mass star forming regions, molecular outflows, and prestellar cores \citep[e.g.][]{Blake1987,Fayolle2015,Bottinelli2004a,Bottinelli2007,Arce2008,Oberg2010,Bacmann2012,Cernicharo2012}.  COMs formed at these early stages of star formation can become incorporated into protoplanetary disks \citep{Visser2009,Visser2011} and further into planetesimals and planets, seeding nascent planets with complex organic material.  COM abundances around protostars are thus of considerable interest for the study of origins of life.  Low-mass stars host most planetary systems, and so the molecular inventories towards low-mass young stellar objects (LYSOs) are most relevant for characterizing potentially habitable environments.  Indeed, our own sun is a low-mass star, and observations of COMs towards LYSOs can inform our understanding of the protostellar phase of our own solar system, as well as the uniqueness of our solar system relative to others like it.

COMs are thought to mainly form within the ice mantles coating interstellar dust grains \citep{Herbst2009}.  First, hydrogenation of atoms or small molecules forms small saturated COMs such as CH$_3$OH, so-called "zeroth-generation" species.  Photolysis or radiolysis can then dissociate small molecules to form radicals \citep[reviewed in][]{Oberg2016}.  During protostellar collapse the cloud material is heated, enabling diffusion and recombination of these small molecules and radicals to form larger complex molecules referred to as "first-generation" species.  Modeling has shown that this becomes efficient above 30K \citep{Garrod2006}.  This mechanism produces the same types of molecules that are commonly observed in YSOs, although some COMs are still underproduced in models compared to observations \citep{Caselli2012}.  Once temperatures reach 100-300K, large molecules desorb and react in the gas phase to form "second-generation" molecules.  Gas-phase reactions may also play a role in first-generation chemistry via the desorption of small molecules, namely CH$_3$OH, followed by reaction in the gas phase rather than on grain-surfaces \citep[e.g.][]{Balucani2015a}.  

Both the ice and gas-phase chemistry scenarios predict a close connection between COMs and the original ice composition.  Ice composition is known to vary between sources, especially CH$_3$OH ice abundances \citep{Oberg2011}.  Furthermore, the chemistry may depend on the stellar radiation field and the evolutionary stage of the source in question.  Observed variations in chemical richness may then signify inherent variations between different sources, or simply that objects are observed at different evolutionary stages, or a combination of both.  Which source of variability dominates will affect predictions on the chemical environment in which planets form during later stages of star formation, and how much this can vary between different sources.

LYSOs are thought to undergo evolution from prestellar cores to protostellar envelopes, which are termed hot corinos once the center is warm enough to sublimate water ice \citep[e.g.][]{Caselli2012}.  COMs have been previously detected towards both envelopes and hot corinos of LYSOs \citep[e.g.][]{Cazaux2003,Bottinelli2004,Bottinelli2007,Oberg2011a}.  \citet{Oberg2014} combined these results from the literature with observations towards 6 young LYSOs and found that COM column densities and abundances span orders of magnitude between different sources.  However, for several reasons it has been difficult to draw firm conclusions on typical abundances and the origin of variability between sources.  These protostars span a range of evolutionary stages, from prestellar cores to evolved protostars with hot envelopes and no ice absorption.  Moreover,the sample size is small ($\sim$14 objects), and objects that were chosen for hosting interesting chemistry may not be representative of the true sample distribution.  

Because of uncertainties in "typical" LYSO COM abundances, it is unclear how LYSOs compare with their massive counterparts (MYSOs) in terms of chemical richness.  It has been claimed both that LYSOs are more enhanced in COMs than MYSOs \citep{Bottinelli2007,Herbst2009} and also that LYSO abundances are comparable or smaller than MYSO abundances \citep{Bisschop2008,Oberg2011a}.  A better estimate of typical COM abundances around LYSOs is required to resolve whether these protostars indeed host distinct chemistries.

To clarify both the characteristic chemistry of LYSOs, as well as to make meaningful comparisons with MYSOs, a homogeneous and unbiased sample is required.  Here we present an extension of the pilot survey in \citet{Oberg2014}, yielding a total of 16 low-mass protostars.  Both the 6 previously observed sources and the 10 new sources are Class 0/I protostars, enabling a direct comparison of the chemistry between sources at a similar evolutionary stage.  

In Section \ref{sec_obs} we describe the observations and data reduction.  In Section \ref{sec_res} we present the results, first deriving column densities and rotational temperatures for the observed COMs.  We derive median values for COM abundances from this sample using survival analysis in order to estimate characteristic abundance frequencies, as well as correlations between different COM species. In Section \ref{modeldesc}, we describe a warm-up model used to simulate chemistry around LYSOs, and present the model results.  In Section \ref{sec_disc} we comment on implications for formation chemistry based on our findings.  We also compare our observational results with the chemical model, and with previous studies of different classes of objects.

\section{Observations}
\label{sec_obs}
Source selection and observation strategy are described in detail in \citet{Graninger2016}.  Briefly, 16 Class 0/I YSOs, identified by their IR spectral indices, were selected from the Spitzer \textit{c2d} ice sample presented in \citet{Boogert2008} based on their location in the northern hemisphere and their ice abundances (Table 1).  The sources were observed with the IRAM 30m telescope using the EMIR 90 GHz receiver and the Fourier Transform Spectrometer (FTS) backend. B1-a, B5 IRS1, L1489 IRS, IRAS 04108+2803, IRAS 03235+3004, and SVS 4-5 were observed June 12–-16, 2013 at 93 -– 101 GHz and 109 –- 117 GHz. All other sources were observed on July 23–-28, 2014 at 92 -– 100 GHz and 108 –- 116 GHz.  All observations had a resolution of 200 kHz.  The telescope beam size ranges from 27" at 92 GHz to 21" at 117 GHz.  Excluding the highest frequency spectral window, the rms values range from 2-7 mK; for each source, the rms around the CH$_3$CN 6$_0$-5$_0$ transition at 110.383 GHz is listed in Table 1 to show the variability in rms between the sources observed.

\begin{deluxetable*}{lccccccccccc} 
	\tabletypesize{\footnotesize}
	\tablecaption{Source information of the complete 16-object c2d embedded protostar sample with ice detections}
	\tablecolumns{11} 
	\tablewidth{0.85\textwidth} 
	\tablehead{\colhead{Source}                                      &
		\colhead{R.A.}                                               & 
		\colhead{Dec}                                                &
		\colhead{Cloud}                                              &
		\colhead{L$_{\rm bol}$}                                      &
		\colhead{M$_{\rm env}$}                                      &
		\colhead{$\alpha_{\rm IR}^{\rm a}$}                          &
		\colhead{N(CH$_3$OH)}                                        &
		\colhead{N$(\rm{H_2O_{(ice)}})^{\rm a}$}                             &
		\colhead{X$_{\rm CH_3OH (ice)}^{\rm b}$}                           &
		\colhead{X$_{\rm NH_3 (ice)}^{\rm b}$}                             &
		\colhead{rms}                                                \\
		\colhead{}                                                   &
		\colhead{(J2000.0)}                                          &
		\colhead{(J2000.0)}                                          &
		\colhead{}                                                   &
		\colhead{L$_\odot$}                                          &
		\colhead{M$_\odot$}                                          &
		\colhead{}                                                   &
		\colhead{10$^{13}$ cm$^{-2}$}                                &
		\colhead{10$^{18}$ cm$^{-2}$}                                &
		\colhead{\% H$_2$O}                                          &
		\colhead{\% H$_2$O}                                          &
		\colhead{(mK)}                                               }
\startdata
B1-a\tablenotemark{c} 	            &03:33:16.67	&31:07:55.1			&Perseus	&1.3\tablenotemark{d}	&2.8\tablenotemark{d}	&1.87	&10.21 [3.24]		&10.39 [2.26]	&$<$1.9		&3.33 [0.98]	& 3.6	\\
B1-c		                        &03:33:17.89	&31:09:31.0	        &Perseus	&3.7\tablenotemark{d}   &17.7\tablenotemark{d}	&2.66	&1.69 [0.51]		&29.55 [5.65]	&$<$7.1		&$<$4.04		& 5.5	\\
B5 IRS 1\tablenotemark{c} 	        &03:47:41.61	&32:51:43.8			&Perseus	&4.7\tablenotemark{d}	&4.2\tablenotemark{d}	&0.78	&1.77 [0.46]		&2.26 [0.28]	&$<$3.7		&$<$2.09		& 7.0	\\
HH 300	                            &04:26:56.30	&24:43:35.3	        &Taurus	    &1.27\tablenotemark{e}  &0.03\tablenotemark{f} 	&0.79	&0.24 [0.10]		&2.59 [0.25]	&$<$6.7		&3.46 [0.90]	& 5.8 	\\
IRAS 03235+3004\tablenotemark{c} 	&03:26:37.45	&30:15:27.9	        &Perseus	&1.9\tablenotemark{d}	&2.4\tablenotemark{d}	&1.44	&1.17 [0.08]		&14.48 [2.26] 	&4.2 [1.2]	&4.71 [1.00]	& 4.2 	\\
IRAS 03245+3002                     & 03:27:39.03   &30:12:59.3         &Perseus	&7.0\tablenotemark{d}   &5.3\tablenotemark{d}	&2.70	&1.54 [0.29]		&39.31 [5.65]	&$<$9.8		&$<$4.40		& 3.7 	\\
IRAS 03254+3050	                    &03:28:34.51	&31:00:51.2	        &Perseus	&--	                    &0.3\tablenotemark{d}	&0.90	&-					&3.66 [0.47]	&$<$4.6		&6.66 [1.37] 	& 3.9	\\
IRAS 03271+3013	                    &03:30:15.16	&30:23:48.8	        &Perseus	&0.8\tablenotemark{d}   &1.2\tablenotemark{d}	&2.06	&0.42 [0.04]		&7.69 [1.76] 	&$<$5.6		&6.37 [1.86]	& 4.8 	\\
IRAS 04108+2803\tablenotemark{c} 	&04:13:54.72	&28:11:32.9	        &Taurus	    &0.62\tablenotemark{e}  &-- 					&0.90	&1.04 [0.44]		&2.87 [0.4]		&$<$3.5		&4.29 [1.03] 	& 4.0 	\\
IRAS 23238+7401	                    &23:25:46.65    &74:17:37.2         &CB244      &--                     &--						&0.95	&2.19 [1.01]		&12.95 [2.26] 	&$<$3.6		&$<$1.24 		& 2.7	\\
L1014 IRS	                        &21:24:07.51    &49:59:09.0         &L1014      &--                     &--						&1.28	&0.88 [0.56]		&7.16 [0.91]	&3.1 [0.8]	&5.20 [1.43] 	& 2.8 	\\
L1448 IRS1                          & 03:25:09.44   & 30:46:21.7        &Perseus	&17.0\tablenotemark{d}  &16.3\tablenotemark{d} 	&0.34	&0.23 [0.04]		&0.47 [0.16]	&$<$14.9	&$<$4.15 		& 3.7	\\ 
L1455 IRS3                          &03:28:00.41	&30:08:01.2	        &Perseus	&0.32\tablenotemark{d}  &0.2\tablenotemark{g}	&0.98	&1.46 [0.88]		&0.92 [0.37]	&$<$12.5	&6.21 [3.51]	& 3.9	\\
L1455 SMM1                          &03:27:43.25	&30:12:28.8         &Perseus	&3.1\tablenotemark{d}   &5.3\tablenotemark{d}	&2.41	&1.48 [0.76]		&18.21 [2.82]	&$<$13.5	&$<$8.29		& 4.1	\\
L1489 IRS\tablenotemark{c}          &04:04:43.07	&26:18:56.4		    &Taurus	    &3.7\tablenotemark{e}   &0.1\tablenotemark{h}	&1.10	&0.69 [0.14]		&4.26 [0.51]	&4.9 [1.5]	&5.42 [0.96] 	& 5.4	\\
SVS 4-5\tablenotemark{c}         	&18:29:57.59	&01:13:00.6			&Serpens	&38\tablenotemark{i}    &-- 					&1.26	&11.19 [4.29]		&5.65 [1.13]	&25.2 [3.5]	&$\sim$4.3	 	& 3.9	\\
\enddata
\tablecomments{Adapted from \citet{Graninger2016}}
\noindent{$^{\rm a}$\citet{Boogert2008}, $^{\rm b}$\citet{Bottinelli2010}, $^{\rm c}$Sources were observed by \citet{Oberg2014}, $^{\rm d}$\citet{Hatchell2007}, $^{\rm e}$\citet{Furlan2008},$^{\rm f}$\citet{Arce2008},$^{\rm g}$\citet{Enoch2009}, $^{\rm h}$\citet{Brinch2007}, $^{\rm i}$\citet{Pontoppidan2004}}

	\label{sources}
\end{deluxetable*}
\clearpage
Spectra were reduced using CLASS\footnote{http://www.iram.fr/IRAMFR/GILDAS}.  Global baselines were fit to each 4 GHz spectral window using several line-free windows.  Each individual scan was baseline subtracted and averaged.  The beam efficiency was modified using Ruze's equation with scaling factor 0.861 and sigma of 63.6 microns, resulting in beam efficiencies at the first, middle, and last channel of 0.8106, 0.7975, and 0.7830.  Together with a forward efficiency of 0.95, the antenna temperature was converted to the main beam temperature T$_{mb}$.  Literature source velocities were used to convert spectra to rest frequency, with fine-tuning adjustments made with the CH$_{3}$OH 2-1 and CN 1-0 ladders.  

\section{Observational Results}
\label{sec_res}
\subsection{Molecule Detections}
\begin{figure*}[h!]
	\centering
	\includegraphics[width=0.8\linewidth]{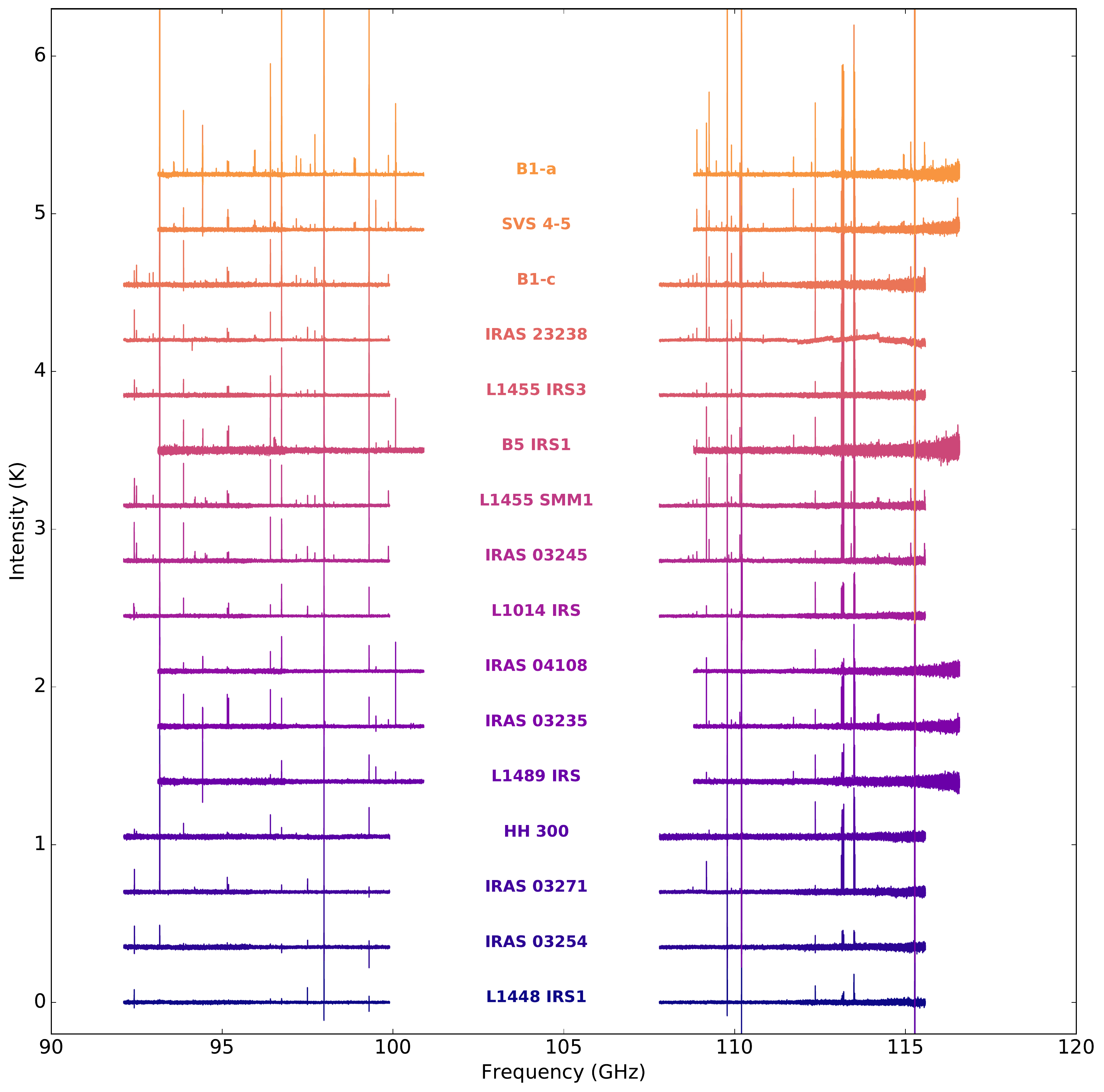}
	\caption{IRAM 30m spectra towards low-mass YSO sample, in order of line richness.}
	\label{specAll}
\end{figure*}

Figure \ref{specAll} shows the spectra of all observed sources.  A wide dispersion in line richness is evident in the sample.  B1-a and SVS 4-5 are very line-dense, followed by a collection of moderately rich sources: B1-c, IRAS 23238, L1455 IRS3, B5 IRS1 L1455 SMM1, IRAS 03235, L1014 IRS, IRAS 04108, and IRAS 03235.  Finally, L1489, HH 300, IRAS 03271, IRAS 03253, and L1448 IRS1 are quite line-poor.
A selection of complex organic molecules that have been observed towards other high- or low-mass protostars are covered by our spectral setting.  We have focused on COMs that were detected towards several of the sources in our survey: CH$_3$CHO, CH$_3$OCH$_3$, CH$_3$OCHO, and CH$_3$CN.  For comparison, we also include the smaller organics HNCO and CH$_2$CO\footnote{The CH$_2$CO transition is only available for sources in the pilot survey: B1-a, B5 IRS1, L1489 IRS, IRAS 04108, IRAS 03235, and SVS 4-5}, as well as the carbon chain cyanide HC$_3$N.  Line candidates within the observed frequency range were identified using the JPL\footnote{http://spec.jpl.nasa.gov} and CDMS\footnote{http://www.astro.uni-koeln.de/cdms/catalog} catalogs, limited by upper excitation energies less than 200 K. 

CH$_2$CO, CH$_3$CHO, CH$_3$OCH$_3$, CH$_3$OCHO, CH$_3$CN, HNCO, and HC$_3$N are detected in 4, 6, 2, 2, 7, 13, and 12 sources respectively; this corresponds to detection percentages of 67\%, 37\%, 13\%, 13\%, 44\%, 81\%, and 75\% respectively.  A molecule is considered to be detected provided that (1) at least one line with a 5$\sigma$ detection or two lines with 3$\sigma$ detections are observed, (2) there is no confusion with common interstellar or YSO molecular lines, and (3) non-detected lines have upper limits that are consistent with the populations predicted by detected lines.  Upper limit treatments are discussed subsequently in more detail.  Because even the line-rich sources in our survey are line-poor in comparison to hot cores, overlapping lines are generally not a concern, and a single line is sufficient to claim a detection when there are no competing line identifications.

\begin{figure}[h!]
	\centering
	\includegraphics[width=0.9\linewidth]{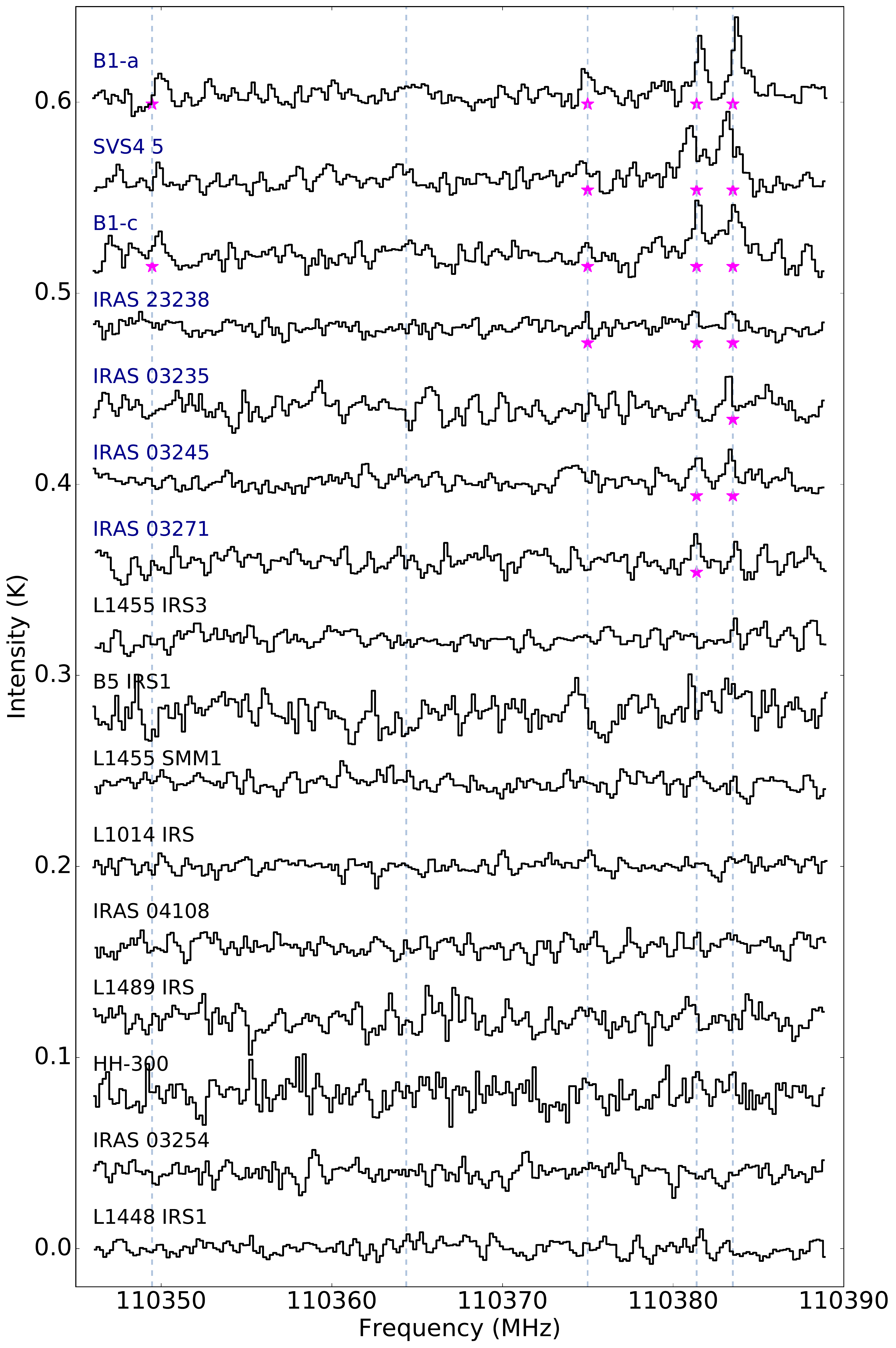}
	\caption{Blow-up of frequency range containing the CH$_3$CN 6-5 ladder.  Grey dashed lines show the line centers, sources with positive detections are written in blue font, and individual detected lines are highlighted with a pink star.}
	\label{specCH3CN}
\end{figure}

Figure \ref{specCH3CN} shows the spectral window containing the CH$_3$CN 6-5 ladder, with detections highlighted with a pink star.  For all other molecules, spectral windows with detections are shown in the Appendix (Figures \ref{spec_ch3cho}-\ref{spec_ch3ocho}).

\subsection{Rotational Diagrams}
Integrated intensities for each observed line were determined by fitting single Gaussians to each feature, and are listed in Tables 6-12 in the Appendix.  For lines with substantial wings, only the central peak was used in deriving column densities.  Unresolved multiplets originating from the same species were treated as a single line by combining the degeneracies and line intensities.  Overlapping lines with contributions from different species were excluded from further analysis.  We assume a calibration uncertainty of 10\%. 

For molecules with multiple line detections in a source we use rotational diagrams to calculate column densities \citep{Goldsmith1999}.  The rotational diagram for CH$_3$CHO is shown in Figure \ref{RDCH3CHO}, with all additional diagrams for detected molecules in the Appendix (Figures \ref{RDCH2CO}-\ref{RDHNCO}).  Tables 2 and 3 summarize the resulting column densities and rotational temperatures.  Rotational diagrams from the pilot survey were refit to include additional lines that were not identified in \citet{Oberg2014}.

Upper level populations for each line are calculated by:
\begin{equation}
	\frac{N_u}{g_u} = \frac{3k\int{T_{mb}dV}}{8\pi^3\nu\mu^2S}.
	\label{N_u}
\end{equation}
Here, $N_u$ is the column density of molecules in the upper level, $g_u$ is the upper level degeneracy, $k$ is Boltzmann's constant, $\int{T_{mb}dV}$ is the integrated main beam temperature, $\nu$ is the transition frequency, $\mu$ is the dipole moment, and $S$ is the transition strength.  Assuming optically thin lines and local thermodynamic equilibrium (LTE), each molecule's total column density $N_{Tot}$ and rotational temperature $T_{rot}$ in each source can be determined from:
\begin{equation}
	\frac{N_u}{g_u} = \frac{N_{Tot}}{Q(T_{rot})} e^{-E_{u}/T_{rot}}
	\label{N_t}
\end{equation}
where $Q(T_{Rot})$ is the partition function and $E_u$ is the energy of the upper level.  We assume that all molecules within the beam can be described by a single temperature \citep[e.g.][]{Bisschop2007a,Oberg2014,Fayolle2015}.

For detected molecules, upper limits are calculated and included in the rotational diagrams for all listed transitions with an equal or greater transition strength and an upper level energy at most 15K greater than the detected lines.  3$\sigma$ upper limits were calculated according to: 
\begin{equation}
\sigma = \mathrm{rms \times FWHM}/\sqrt{n_{ch}}
\label{3sig}
\end{equation}
Here, the rms is taken from a 40 km s$^{-1}$ spectral window containing the transition and FWHM is equal to the average FWHM of detected lines for that molecule in the same source.  $n_{ch}$ is the number of channels across the FWHM, in this case equal to $\sim$FWHM/0.6 km  s$^{-1}$ channel$^{-1}$.  Equation \ref{N_u} is used to calculate the population upper limits.

In cases where a single transition was detected, the column densities were calculated adopting the average rotational temperatures in the sample for the species in question, if available. For HNCO and CH$_2$CO, at most one transition was detected for all sources.  The average rotational temperature of HC$_3$N in our sample (14K) was used for calculating HNCO column densities and that of CH$_3$CHO (8K) used for calculating CH$_2$CO column densities.  This is based on spatial trends in emission observed in high-mass protostars \citep{Bisschop2007a,Oberg2013,Fayolle2015}. 

For molecules with no detected transitions in a given source, a 3$\sigma$ upper limit was calculated using equation \ref{3sig}.  In this case, the rms for each non-detected molecule was taken from a 40 km s$^{-1}$ spectral window containing its lowest-energy transition.  We assume a FWHM equal to the average FWHM of all detected lines within a given source; for sources with no detections (HH-300 and L1448 IRS1) we assume the FWHM of CH$_3$OH within the source.  Upper limits for total column densities were calculated from equations \ref{N_u} and \ref{N_t} assuming the sample-averaged rotational temperatures.

\begin{figure}[]
	\centering
	\includegraphics[width=0.9\linewidth]{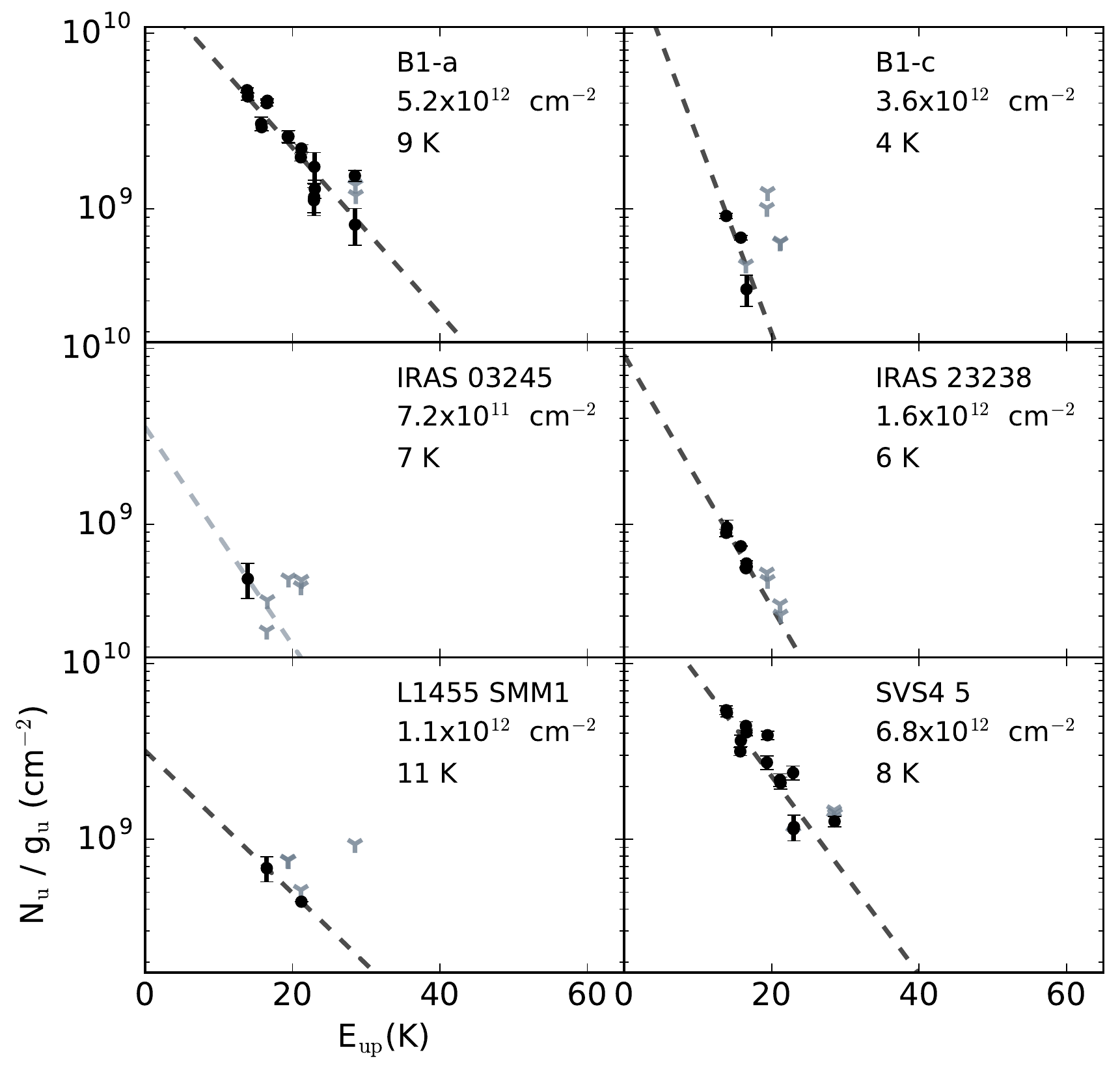}
	\caption{Rotation diagrams for CH$_{3}$CHO.  Black circles indicate detections and grey triangles indicate upper limits.  Black dashed lines represent the fits to the data.  When a line could not be fit, a rotational temperature was assumed as described in the text, shown in grey dashed lines.}
	\label{RDCH3CHO}
\end{figure}

We note that B1-a and B1-c appear to have a detection of CH$_3$CN 6$_5$-6$_4$ at 110.349 GHz, which has an upper energy of 132.8K.  We exclude these points from fitting the rotational diagram.  The peak velocity of this transition is slightly offset from the other CH$_3$CN detections for both sources in which we observe it; thus, if real this CH$_3$CN line does not originate from the same location as the other lines.  The line may trace CH$_3$CN in a shock, but it is more likely due to an unidentified carrier; the latter was previously suggested by \citet{Bottinelli2007}, who observed an excess of this CH$_3$CN transition towards the hot corino NGC1333-IRAS4B.

\begin{deluxetable*}{lllllllllllllll} 
	\label{col_dens_1}
	\tabletypesize{\footnotesize}
	\tablecaption{Column densities and rotational temperatures: Oxygen-bearing molecules}
	\tablecolumns{8} 
	\tablewidth{0.85\textwidth} 
	\tablehead{\colhead{}                                           &
		\multicolumn{2}{c}{CH$_{2}$CO}                             &
		\multicolumn{2}{c}{CH$_{3}$CHO}                            &
		\multicolumn{2}{c}{CH$_3$OCH$_3$}                           &
		\multicolumn{2}{c}{CH$_{3}$OCHO}                           \\
		\colhead{}                                                  &
		\colhead{N$_{Tot}$ (cm$^{-2}$)}                             &
		\colhead{T$_{Rot}$ (K)}                                     &
		\colhead{N$_{Tot}$(cm$^{-2}$)}                              &
		\colhead{T$_{Rot}$ (K)}                                     &
		\colhead{N$_{Tot}$ (cm$^{-2}$)}                             &
		\colhead{T$_{Rot}$ (K)}                                     &
		\colhead{N$_{Tot}$ (cm$^{-2}$)}                             &
		\colhead{T$_{Rot}$ (K)}                                     }
	\startdata
	B1-a &8.6 (11.8) x10$^{12}$ &\textit{8 (2)} &5.2 (0.8) x10$^{12}$ & 9 (1)&8.1 (1.3) x10$^{12}$ &22 (5)&5.9 (6.2) x10$^{12}$ &17 (11)\\
	B1-c &- & - &3.6 (3.4) x10$^{12}$ & 4 (1)&$<$1.0x10$^{13}$ &\textit{17 (5)} &$<$9.5x10$^{12}$ &\textit{16 (1)} \\
	B5 IRS1 &$<$1.4x10$^{12}$ &\textit{8 (2)} &$<$1.1x10$^{12}$ &\textit{8 (2)} &$<$1.0x10$^{13}$ &\textit{17 (5)} &$<$9.4x10$^{12}$ &\textit{16 (1)} \\
	HH-300 &- & - &$<$4.5x10$^{11}$ &\textit{8 (2)} &$<$7.1x10$^{12}$ &\textit{17 (5)} &$<$6.5x10$^{12}$ &\textit{16 (1)} \\
	IRAS 03235 &1.0 (1.4) x10$^{12}$ &\textit{8 (2)} &$<$6.1x10$^{11}$ &\textit{8 (2)} &$<$6.4x10$^{12}$ &\textit{17 (5)} &$<$4.1x10$^{12}$ &\textit{16 (1)} \\
	IRAS 03245 &- & - &6.4 (4.9) x10$^{11}$ &\textit{8 (2)} &$<$5.6x10$^{12}$ &\textit{17 (5)} &$<$5.8x10$^{12}$ &\textit{16 (1)} \\
	IRAS 03254 &- & - &$<$6.4x10$^{11}$ &\textit{8 (2)} &$<$8.4x10$^{12}$ &\textit{17 (5)} &$<$5.3x10$^{12}$ &\textit{16 (1)} \\
	IRAS 03271 &- & - &$<$4.8x10$^{11}$ &\textit{8 (2)} &$<$8.5x10$^{12}$ &\textit{17 (5)} &$<$4.9x10$^{12}$ &\textit{16 (1)} \\
	IRAS 04108 &1.4 (1.9) x10$^{12}$ &\textit{8 (2)} &$<$7.6x10$^{11}$ &\textit{8 (2)} &$<$7.2x10$^{12}$ &\textit{17 (5)} &$<$6.0x10$^{12}$ &\textit{16 (1)} \\
	IRAS 23238 &- & - &1.6 (1.0) x10$^{12}$ & 6 (2)&$<$5.1x10$^{12}$ &\textit{17 (5)} &$<$3.7x10$^{12}$ &\textit{16 (1)} \\
	L1014 IRS &- & - &$<$3.1x10$^{11}$ &\textit{8 (2)} &$<$3.9x10$^{12}$ &\textit{17 (5)} &$<$3.2x10$^{12}$ &\textit{16 (1)} \\
	L1448 IRS1 &- & - &$<$4.2x10$^{11}$ &\textit{8 (2)} &$<$6.1x10$^{12}$ &\textit{17 (5)} &$<$4.1x10$^{12}$ &\textit{16 (1)} \\
	L1455 IRS3 &- & - &$<$4.3x10$^{11}$ &\textit{8 (2)} &$<$8.1x10$^{12}$ &\textit{17 (5)} &$<$4.8x10$^{12}$ &\textit{16 (1)} \\
	L1455 SMM1 &- & - &1.1 (0.3) x10$^{12}$ &11 (7)&$<$6.7x10$^{12}$ &\textit{17 (5)} &$<$4.7x10$^{12}$ &\textit{16 (1)} \\
	L1489 IRS &$<$1.4x10$^{12}$ &\textit{8 (2)} &$<$8.7x10$^{11}$ &\textit{8 (2)} &$<$8.1x10$^{12}$ &\textit{17 (5)} &$<$6.4x10$^{12}$ &\textit{16 (1)} \\
	SVS4 5 &9.0 (12.4) x10$^{12}$ &\textit{8 (2)} &6.8 (1.1) x10$^{12}$ & 8 (1)&1.2 (0.5) x10$^{13}$ &12 (3)&7.0 (4.4) x10$^{12}$ &15 (5)\\
		\enddata
	\tablecomments{Uncertainties are listed in parentheses.  T$_{Rot}$ values are calculated from the rotational diagram method when possible; otherwise, sample-averaged values for that molecule are assumed, shown in italics in the table.}
\end{deluxetable*}

\begin{deluxetable*}{lllllllllllll} 
	\label{col_dens_2}
	\tabletypesize{\footnotesize}
	\tablecaption{Column densities and rotational temperatures: Nitrogen-bearing molecules}
	\tablecolumns{8} 
	\tablewidth{0.85\textwidth} 
	\tablehead{\colhead{}                                           &
		\multicolumn{2}{c}{CH$_{3}$CN}                             &
		\multicolumn{2}{c}{HC$_3$N}                                 &
		\multicolumn{2}{c}{HNCO}                                    \\
		\colhead{}                                                  &
		\colhead{N$_{Tot}$ (cm$^{-2}$)}                             &
		\colhead{T$_{Rot}$ (K)}                                     &
		\colhead{N$_{Tot}$(cm$^{-2}$)}                              &
		\colhead{T$_{Rot}$ (K)}                                     &
		\colhead{N$_{Tot}$ (cm$^{-2}$)}                             &
		\colhead{T$_{Rot}$ (K)}                                     }
	\startdata
B1-a &4.9 (1.1) x10$^{11}$ &33 (9)&4.2 (1.2) x10$^{12}$ &12 (2)&7.7 (4.3) x10$^{12}$ &\textit{14 (3)} \\
B1-c &3.5 (0.6) x10$^{11}$ &18 (2)&4.2 (3.4) x10$^{12}$ &\textit{14 (3)} &8.8 (4.9) x10$^{12}$ &\textit{14 (3)} \\
B5 IRS1 &$<$1.7x10$^{11}$ &\textit{27 (7)} &3.2 (1.2) x10$^{12}$ &11 (2)&3.5 (2.0) x10$^{12}$ &\textit{14 (3)} \\
HH-300 &$<$1.2x10$^{11}$ &\textit{27 (7)} &$<$1.3x10$^{11}$ &\textit{14 (3)} &$<$6.6x10$^{11}$ &\textit{14 (3)} \\
IRAS 03235 &1.6 (0.9) x10$^{11}$ &\textit{27 (7)} &3.9 (1.0) x10$^{12}$ &13 (2)&1.8 (1.0) x10$^{12}$ &\textit{14 (3)} \\
IRAS 03245 &1.9 (1.5) x10$^{11}$ &33 (28)&3.6 (2.9) x10$^{12}$ &\textit{14 (3)} &2.8 (1.6) x10$^{12}$ &\textit{14 (3)} \\
IRAS 03254 &$<$1.0x10$^{11}$ &\textit{27 (7)} &$<$9.2x10$^{10}$ &\textit{14 (3)} &$<$6.1x10$^{11}$ &\textit{14 (3)} \\
IRAS 03271 &1.9 (1.1) x10$^{11}$ &\textit{27 (7)} &1.7 (1.4) x10$^{12}$ &\textit{14 (3)} &1.4 (0.8) x10$^{12}$ &\textit{14 (3)} \\
IRAS 04108 &$<$1.2x10$^{11}$ &\textit{27 (7)} &$<$1.0x10$^{11}$ &\textit{14 (3)} &7.0 (4.0) x10$^{11}$ &\textit{14 (3)} \\
IRAS 23238 &1.4 (0.3) x10$^{11}$ &33 (6)&3.2 (2.6) x10$^{12}$ &\textit{14 (3)} &4.6 (2.6) x10$^{12}$ &\textit{14 (3)} \\
L1014 IRS &$<$6.2x10$^{10}$ &\textit{27 (7)} &4.4 (3.6) x10$^{11}$ &\textit{14 (3)} &1.7 (0.9) x10$^{12}$ &\textit{14 (3)} \\
L1448 IRS1 &$<$9.4x10$^{10}$ &\textit{27 (7)} &$<$8.7x10$^{10}$ &\textit{14 (3)} &$<$4.4x10$^{11}$ &\textit{14 (3)} \\
L1455 IRS3 &$<$9.1x10$^{10}$ &\textit{27 (7)} &6.5 (5.3) x10$^{11}$ &\textit{14 (3)} &1.7 (0.9) x10$^{12}$ &\textit{14 (3)} \\
L1455 SMM1 &$<$1.1x10$^{11}$ &\textit{27 (7)} &2.4 (1.9) x10$^{12}$ &\textit{14 (3)} &2.4 (1.3) x10$^{12}$ &\textit{14 (3)} \\
L1489 IRS &$<$1.4x10$^{11}$ &\textit{27 (7)} &3.5 (2.4) x10$^{11}$ &20 (13)&7.7 (4.7) x10$^{11}$ &\textit{14 (3)} \\
SVS4 5 &5.2 (0.9) x10$^{11}$ &17 (2)&1.1 (0.3) x10$^{13}$ &13 (2)&7.0 (3.9) x10$^{12}$ &\textit{14 (3)} \\
	\enddata
	\tablecomments{Uncertainties are listed in parentheses.  T$_{Rot}$ values are calculated from the rotational diagram method when possible; otherwise, sample-averaged values for that molecule are assumed, shown in italics in the table.}
\end{deluxetable*}

\subsection{Column densities and abundances}
\subsubsection{COM detections and upper limits}
\begin{figure*} 
	\centering
	\includegraphics[width=0.95\linewidth]{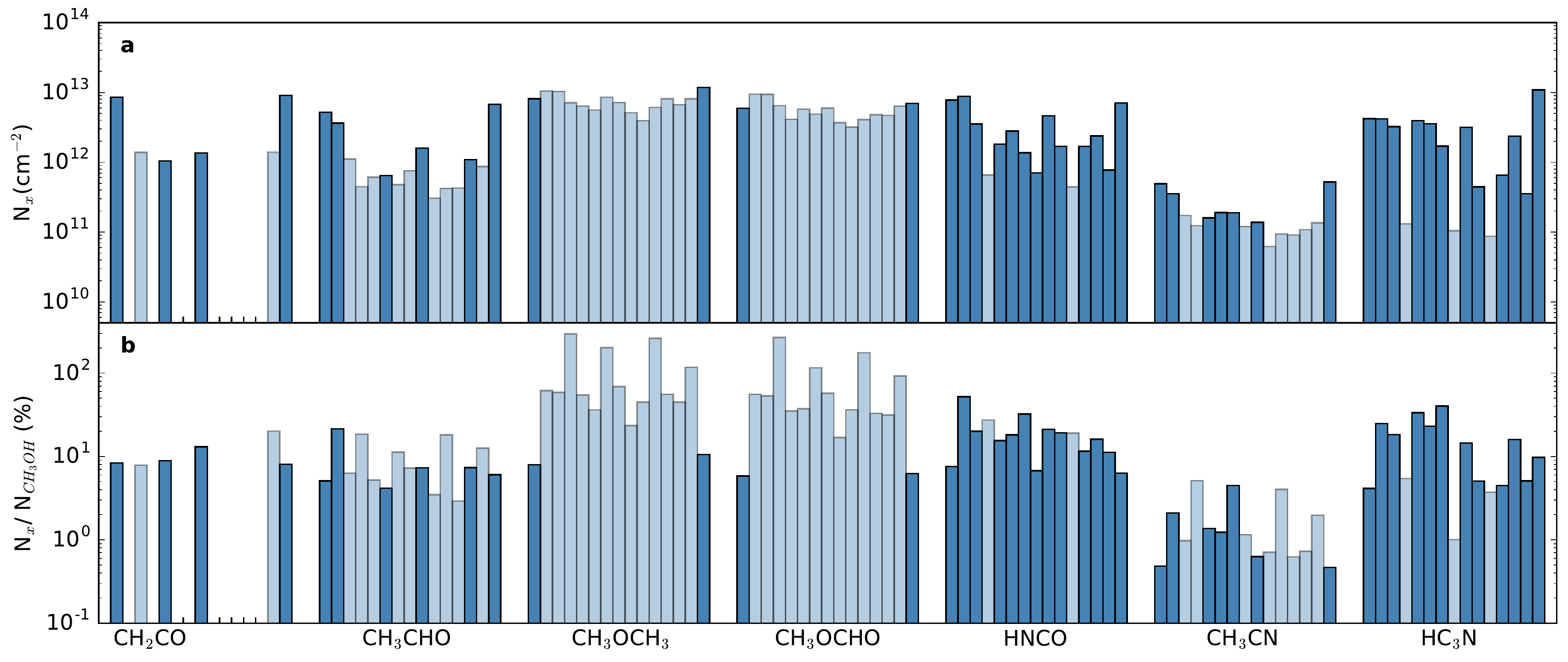}
	\caption{a: Observed column densities for each molecule.  b: Abundances with respect to CH$_3$OH.  For both panels, detections are shown in dark blue, and upper limits in light blue.  For each molecule, sources are ordered alphabetically: B1-a, B1-c, B5 IRS1, HH-300, IRAS 03235, IRAS 03245, IRAS 03271, IRAS 04108, IRAS 23238, L1014 IRS, L1448 IRS1, L1455 IRS3, L1455 SMM1, L1489 IRS, SVS 4-5.}
	\label{dist_dets}
\end{figure*}

Figure \ref{dist_dets}a shows the column densities derived for each molecule in all sources, both detections and upper limits.  CH$_3$OCH$_3$ and CH$_3$OCHO are each detected in only two sources.  These detections are close to the upper limits for sources without detections.  Thus, the upper limits for CH$_3$OCH$_3$ and CH$_3$OCHO are not very constraining, and the variability in column density of these molecules cannot be well understood from this sample.  All other molecules have multiple detections that are substantially larger than the non-detection upper limits, and the resulting ranges of derived column densities span at least an order of magnitude.  

The sources B1-a and SVS 4-5, shown as the first and last bar in each grouping in Figure \ref{dist_dets}, stand out as being quite enhanced relative to the other sources.  All 7 molecules are detected towards these sources, and the derived column densities are high compared to the sample medians.  In contrast, the sources L1448 IRS1 and HH-300 have no detections.  The majority of sources exhibit a range of moderate column densities, with a few but not all molecules detected in each source.

Abundances with respect to CH$_3$OH are shown in Figure \ref{dist_dets}b. CH$_3$OH column densities are taken from \citet{Graninger2016}.  Because column densities of CH$_3$OH are not available for IRAS 03254, we omit this source from comparison.  Once normalized to CH$_3$OH, B1-a and SVS 4-5 molecule abundances are consistent or under-abundant compared with other sources, suggesting that their high column densities do not signify a distinct chemsitry.

\subsubsection{Median column densities and abundances}
Median abundances were first calculated using detections only; however, this neglects the information provided by upper limits, and thus risks over-estimating the median occurrences.  Survival analysis using the Kaplan-Meier (KM) estimate of the survival function with left censorship was performed in order to account for this.  In this method, all detections and non-detections are ordered, and then the values of positive detections are used to divide the total range of values into intervals.  Upper limits within an interval are counted as having the lower delimiting value of the interval.  In other words, each positive detection is weighted by the number of upper limits that occur between it and the next largest positive detection.  Further details can be found in \citet{Feigelson1985} and \citet{Miller1981}.

\begin{deluxetable*}{lccccccc}
	\label{tab_median}
	\tabletypesize{\footnotesize}
	\tablecaption{Median column densities and abundances}
	\tablecolumns{8} 
	\tablewidth{0.85\textwidth} 
	\tablehead{\colhead{}                                      &
		\multicolumn{2}{c}{Median column densities}            & 
		\colhead{Median abundances}                            \\ 
		\colhead{}                                             &  
		\multicolumn{2}{c}{(10$^{12}$ cm$^{-2}$)}              &
		\colhead{(\% with respect to CH$_3$OH)}                \\
		\colhead{}											   &
		\colhead{Detections only}                              & 
		\colhead{Survival analysis\tablenotemark{a}}           & 
		\colhead{Survival analysis\tablenotemark{a}}           }
	\startdata
	CH$_2$CO       & 5.0   & 1.2 $_{0.78}^{5.0}$    & 8.2 $_{7.9}^{8.8}$  \\
	CH$_3$CHO      & 2.6   & 0.27 $_{0.14}^{1.2}$  & 4.4 $_{2.3}^{6.1}$  \\
	CH$_3$OCH$_3$  & 9.9   & 2.4 $_{1.2}^{3.5}$    & 5.3 $_{4.0}^{9.2}$  \\
	CH$_3$OCHO     & 6.5   & 1.9 $_{1.0}^{2.9}$    & 3.1 $_{2.9}^{6.0}$  \\
	HNCO           & 2.4   & 1.8 $_{0.76}^{3.8}$   & 16 $_{7.9}^{20.}$  \\
	CH$_3$CN       & 0.19  & 0.060 $_{0.03}^{0.19}$ & 0.48 $_{0.46}^{1.1}$  \\
	HC$_3$N        & 3.2   & 2.0 $_{0.29}^{3.7}$   & 7.4 $_{4.2}^{20.}$
	\enddata
	\tablenotetext{$\rm{a}$}{Lower and upper quartiles shown to the right of the median}
\end{deluxetable*}

\begin{figure}[] 
	\centering
	\includegraphics[width=0.7\linewidth]{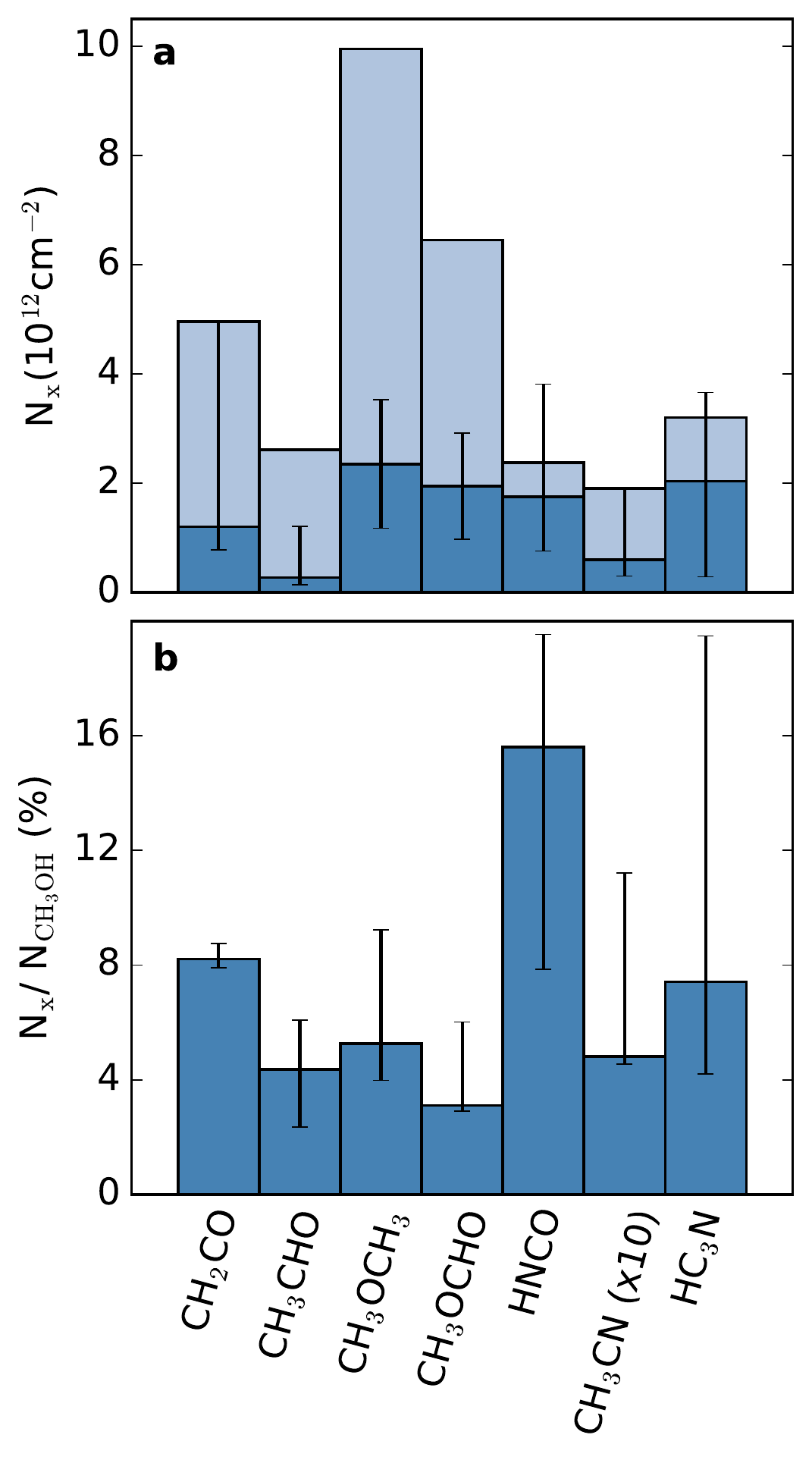}
	\caption{a: Median column densities of each molecule, calculated using detections only (light blue) and using survival analysis (dark blue). b: Abundances for each molecule calculated with survival analysis.  In both panels, CH$_3$CN has been scaled by 10 for clarity.  Error bars span the first and third quartile determined using survival analysis.}
	\label{fig_medians}
\end{figure}

Since the survivial function is discrete, median abundances are calculated by linear interpolation between the values above and below where the cumulative density function (CDF) is equal to 0.5.  However, medians cannot be computed by the KM estimate for samples with only upper limits in the lowest 50\% of values since the first positive detection occurs after the cumulative density has already exceeded 0.5.  In this sample, this applies to CH$_{3}$OCH$_{3}$, CH$_{3}$OCHO, and CH$_{3}$CN.  To mitigate this we calculate medians using the KM estimate with the lowest value assigned a "detection" status, regardless of its true identity as a detection or a non-detection. This may result in slightly elevated estimates for median values, but as seen in Figure \ref{fig_medians} it is still a more realistic estimate of the median than using detections only. 

Median column densities for each molecule are shown in Figure \ref{fig_medians}a and Table 4 for both methods.  There is a clear difference between the medians calculated from only detections and those using survival analysis, demonstrating the importance of using the constraints provided by non-detections.  

Median column densities of all species except CH$_3$CN are on the order of 10$^{12}$ cm$^{-2}$, while CH$_3$CN is over an order of magnitude lower. CH$_3$OCHO and CH$_3$OCH$_3$ have among the highest median column densities compared to the other observed molecules, but this may be somewhat misleading since each is detected in only two sources.  Given the large values of the partition function for these molecules, detections are possible only for sources with very high abundances and thus the statistics are biased, even with the inclusion of upper limits.  HC$_3$N and HNCO also have fairly high column densities and are detected in 12 and 13 sources respectively, indicating that they are both common and abundant in low-mass protostellar environments.

Median abundances using survival analysis are shown in Figure \ref{fig_medians}b.  Most molecules have median abundances of a few percent with respect to CH$_3$OH; CH$_3$CN is again the exception, with a median abundance an order of magnitude lower.  The distribution of abundances, reflected in the error bars spanning the lower and upper quartile, is much tighter for oxygen-bearing molecules than for nitrogen-bearing molecules; this is discussed further in the following section.  

\subsubsection{Column density and abundances distributions}
\label{sec_dists}
\begin{figure*} 
	\centering
	\includegraphics[width=0.95\linewidth]{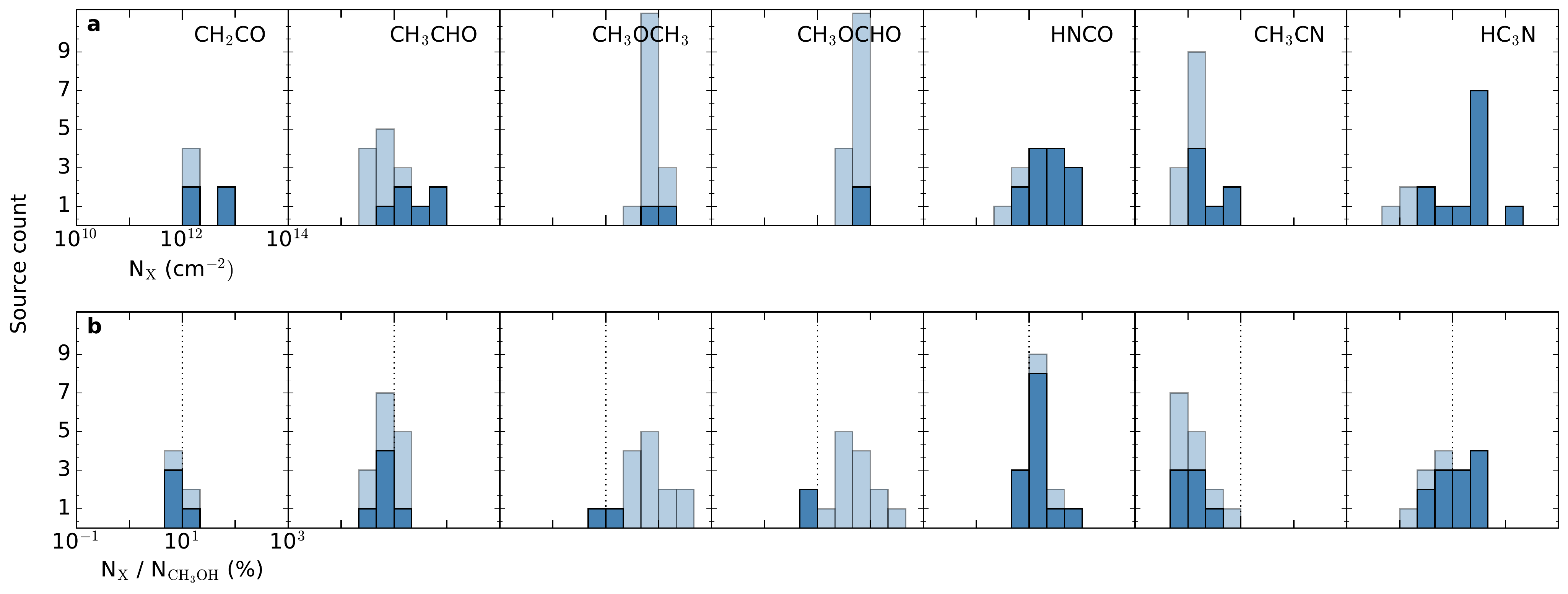}
	\caption{a: Histograms of observed column densities for each molecule.  b: Abundances with respect to CH$_3$OH.  The dotted black line represents 10\%.  For both panels, detections are shown in dark blue, and upper limits in light blue.}
	\label{hist_dets}
\end{figure*}

To explore the distributions of column densities before and after normalization to CH$_3$OH, Figure \ref{hist_dets}a shows histograms of each molecule's column density.  For most molecules, the range of column densities spans an order of magnitude or more.  The exceptions are CH$_3$OCH$_3$ and CH$_3$OCHO, for which only two detections of each molecule are available.  The lower and upper quartiles for column densities are listed in Table 4.  HC$_3$N has a particularly broad distribution, with over an order of magnitude difference between the lower and upper quartile values.  CH$_2$CO, CH$_3$CHO, and CH$_3$CN have fairly large spreads, with a factor of $\sim$6-7 difference in lower and upper quartile values.  The narrow distributions of CH$_3$OCH$_3$ and CH$_3$OCHO are not informative due to the small number of detections.

When normalized to CH$_3$OH, the distribution of abundances of oxygen-bearing species tightens considerably as seen in the narrower distributions in Figure \ref{hist_dets}b compared to \ref{hist_dets}a, with most detections around 10\% with respect to CH$_3$OH.  In contrast, the CH$_3$OH-normalized abundance distributions for nitrogen-bearing species are broad.  The lower and upper quartiles in Table 4 reflect this as well: for the abundances of oxygen-bearing species, upper quartile values are 1-2 times larger than lower quartile values.  In contrast, upper quartile values for nitrogen-bearing species are 3-4 times larger than lower quartile values. Together, this suggests an oxygen chemistry that is connected to CH$_3$OH, whereas the nitrogen chemistry is regulated by other factors besides CH$_3$OH abundances.

\subsection{COM Rotational temperatures}
For sources with at least 3 detections of a given molecule, rotational excitation temperatures could be derived using the rotational diagram method.  These spanned from 4K to 33K, with CH$_3$CHO having the lowest temperature and CH$_3$CN the highest.  All rotational temperatures are quite low, indicating that the majority of emission does not come from a region that is both hot and dense.  Both the temperature and density of protostars increase towards the center, therefore most emission must originate on larger scales.  Since rotational temperatures depend on the excitation conditions and each molecule's excitation properties, they are typically not equal to the kinetic temperature.  However, for molecules with similar dipole moments, rotational temperatures should be proportional to kinetic temperatures; for this reason we can compare CH$_3$CHO and CH$_3$CN with CH$_3$OH to put qualitative constraints on their emission regions.  As for previous observations of low-mass \citep{Oberg2014} and high-mass \citep{Bisschop2007a,Fayolle2015} protostars, we find a relatively low rotational temperature for CH$_3$CHO, suggesting that it is mainly emitting from the cold outer envelope.  Likewise, we find a higher rotational temperature for CH$_3$CN, suggesting that it emits from closer in.

The rotational temperatures of CH$_3$OCH$_3$ and CH$_3$OCHO are both higher than the CH$_3$OH rotational temperature in the two sources in which they are detected (B1-a and SVS 4-5).  Given this, it is unlikely that the envelope is the sole origin of emission.  This is consistent with observations of hot-core MYSOs in which both molecules are classified as hot molecules \citep{Bisschop2007a}.  We note, however, that CH$_3$OCH$_3$ and CH$_3$OCHO have low signal-to-noise detections and are found only in two sources, and so further observations are required to confirm whether these molecules are indeed warm emitters in LYSOs.

\subsection{Correlation studies}
\label{sec_PCC}
To explore chemical relationships between different COMs, we determined the correlations between each COM with one another and with CH$_3$OH.  Strong positive correlations are expected for chemically related molecules (e.g. if one forms from another), or for molecules that depend similarly on an underlying variable such as envelope mass or temperature.  When correlating column densities, some correlation is always expected since all molecules typically increase with an increasing total column density in a line of sight.  This could in theory be divided out, but there are insufficient constraints on the physical characteristics of each source and on the origin of COM emission within each source to convert column densities to abundances with respect to H$_2$.  We therefore use the column density correlation strengths to infer which molecules are more closely related, with the caveat that some of the correlation may be due to source richness rather than chemical relatedness.

\begin{figure*} 
	\centering
	\includegraphics[width=0.8\linewidth]{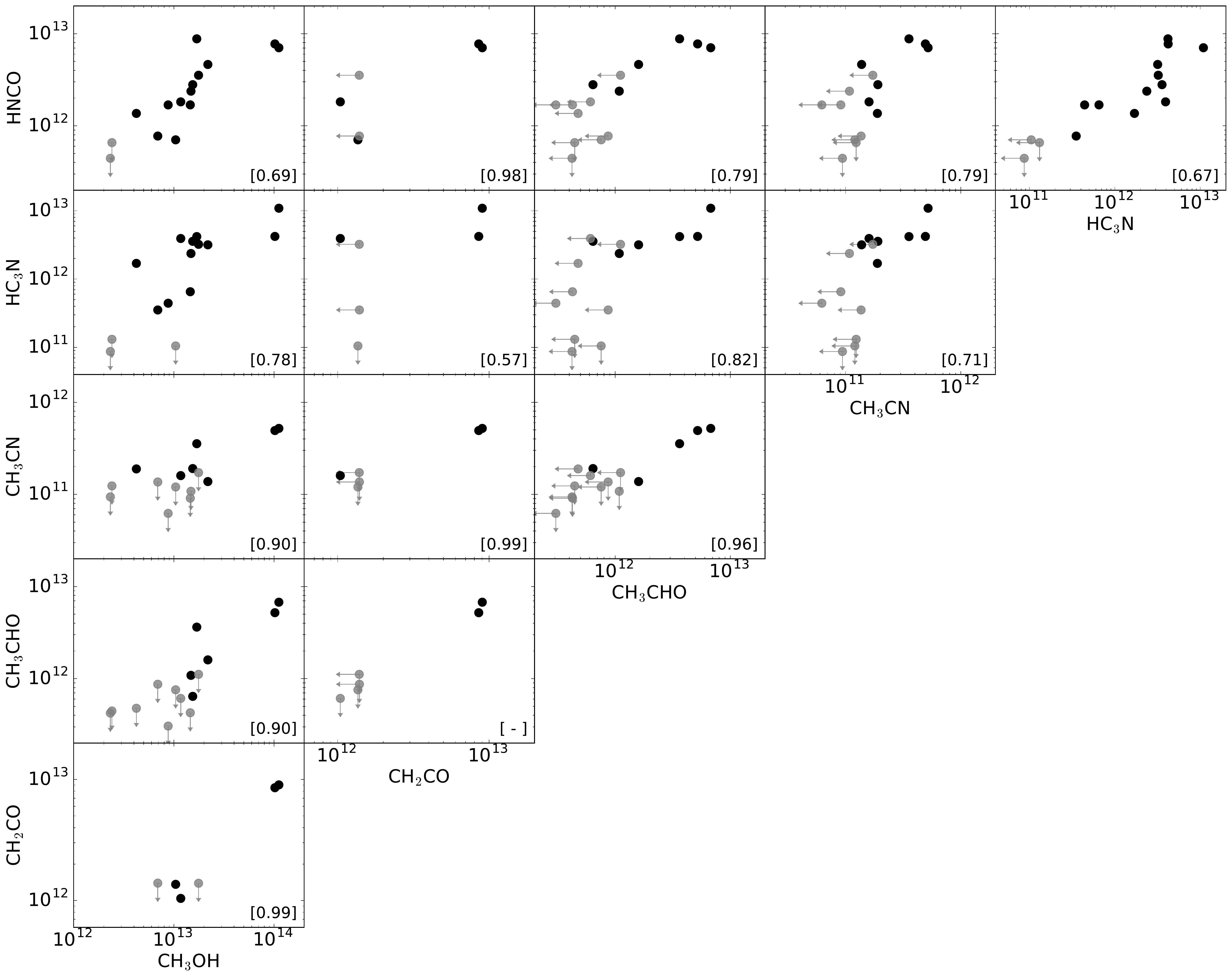}
	\caption{Column densities (cm$^{-2}$) of each molecule plotted against one another.  Detections are shown as black circles and upper limits as grey arrows.  Pearson correlation coefficients are listed in brackets.}
	\label{fig_ch3ohcorr}
\end{figure*}

We calculate Pearson correlation coefficients for each COM with one another, shown in Table 5.  Only detections are used for this calculation; because of this, only pairs of molecules which are detected in at least three of the same sources have correlation coefficients listed.  Figure \ref{fig_ch3ohcorr} shows scatterplots of each molecule's column density plotted against one another.

\begin{deluxetable*}{llllll} 
	\tabletypesize{\footnotesize}
	\tablecaption{Column density correlations between molecules.}
	\tablecolumns{6} 
	\tablewidth{0.8\linewidth} 
	\tablehead{\colhead{}                                 &
		\colhead{CH$_3$OH}                                    &
		\colhead{CH$_2$CO}                                 &
		\colhead{CH$_3$CHO}                             &
		\colhead{CH$_3$CN}                            &
		\colhead{HC$_3$N}                                }
	
	\startdata
	HNCO       &0.69 [13]  &0.98 [4]  &0.79 [6]  &0.79 [7] &0.67 [12] \\
	HC$_3$N    &0.78 [12]  &0.57 [3]  &0.82 [6]  &0.71 [7] \\
	CH$_3$CN   &0.90 [7]   &0.99 [3]  &0.96 [5]  \\
	CH$_3$CHO  &0.90 [6]   & -   \\
	CH$_2$CO   &0.99 [4]  \\
	\enddata
	\tablecomments{Brackets indicate the number of sources with detections for both molecules.  A dash indicates a pair of molecules with fewer than 3 sources in common.}
	\label{pcc}
\end{deluxetable*}

CH$_3$OH is well-correlated with CH$_2$CO, CH$_3$CHO, and CH$_3$CN, and more weakly correlated with HNCO and HC$_3$N.  The higher correlation of CH$_3$OH with oxygen-bearing species compared to nitrogen-bearing species supports a scenario in which O-bearing COMs form from CH$_3$OH processing.  Within the nitrogen-bearing family, there are no particularly strong relationships between any of the molecules.  HC$_3$N is the least well-correlated with other molecules; this is unsurprising given that it is a carbon chain rather than a hydrogen-rich COM.  CH$_3$OCH$_3$ and CH$_3$OCHO correlations cannot be assessed since each is detected in only two sources.

We caution that the  high correlation values for CH$_2$CO with CH$_3$OH, HNCO and CH$_3$CN may be an artifact of the low number of CH$_2$CO detections: the CH$_2$CO transition at 100.095 GHz is only observable within the spectral range of the 6 objects in the pilot survey, of which it is detected in 4.  The lack of observational constraints for intermediate values of CH$_2$CO hinders an interpretation of the true distribution of column densities.

Correlations of column densities with physical properties may also provide insight into the chemistry.  To this end, we checked for correlations between envelope mass and bolometric luminosities (see Table 1) with our COM column densities. 

\begin{figure}[h!] 
	\centering
	\includegraphics[width=\linewidth]{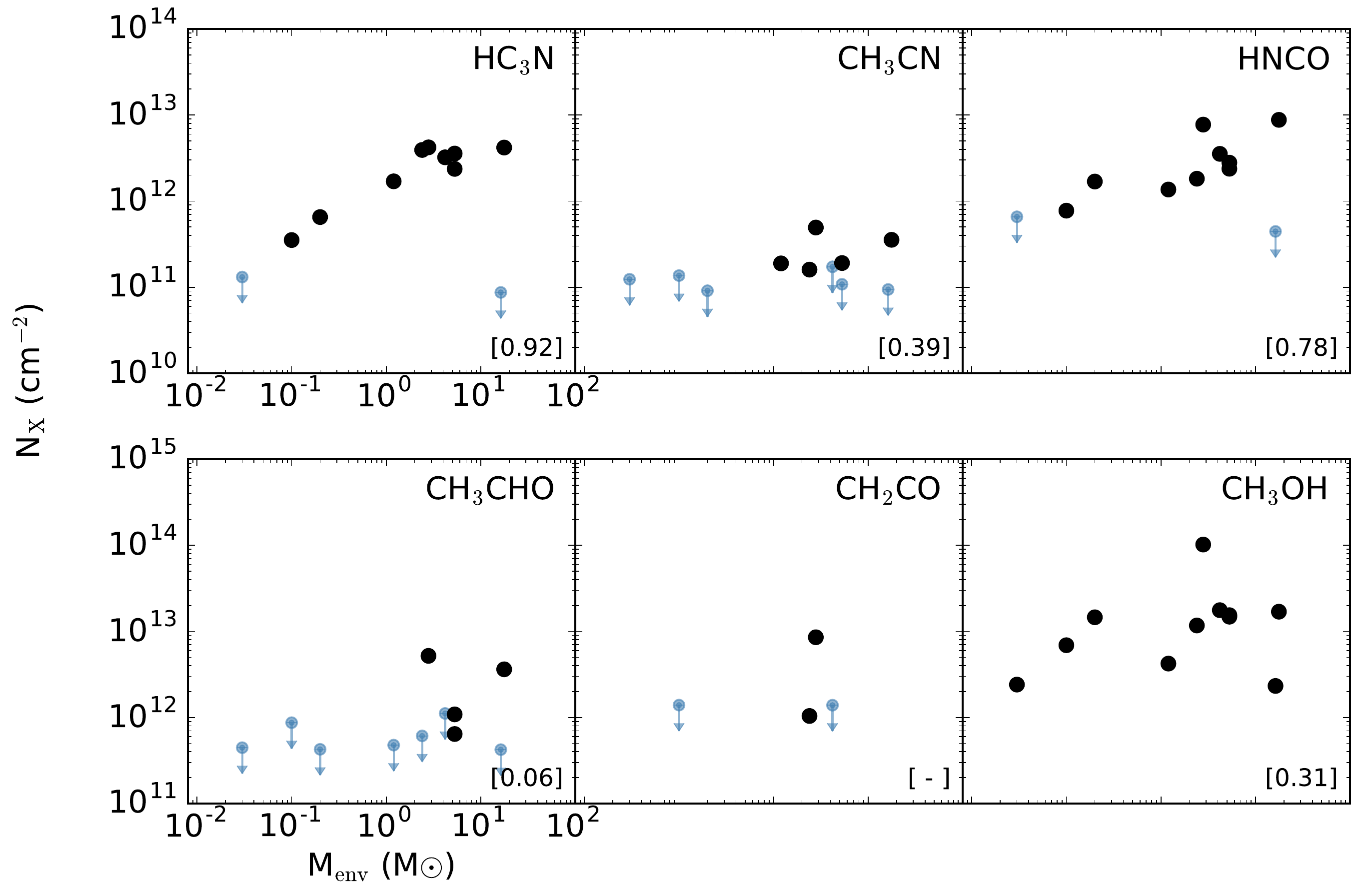}
	\caption{Column densities plotted against envelope mass. Upper limits due to molecule non-detections are shown in blue.  Pearson correlation coefficients are shown in brackets in the lower right corner.}
	\label{Menv}
\end{figure}

All molecules correlate positively with envelope mass (Figure \ref{Menv}).  This is in part because more molecules are present in lines of sight containing more material.  The correlation is strongest for HC$_3$N followed by HNCO, with all other molecules showing weaker positive correlations.  Among the nitrogen-bearing species, the observed sequence of correlation strengths with M$_{\rm{env}}$ are consistent with the expected origin of emission of each molecule.  HC$_3$N is thought to form through gas-phase chemistry and emit from the cold envelope, and therefore should trace the envelope mass very well.  HNCO is thought to emit from both the warm inner region and the cold envelope, which is consistent with a weaker correlation with envelope mass.  Based on its rotational temperature, CH$_3$CN is mainly present in the warmer, more compact region of the protostar, and should therefore not depend strongly on total envelope mass.  

Interestingly, CH$_3$CHO and CH$_2$CO are also cold emitters, but unlike HC$_3$N they depend only weakly on envelope mass.  This could be because these molecules form mainly on grain surfaces rather than via gas-phase chemistry.  

\begin{figure}[h!] 
	\centering
	\includegraphics[width=\linewidth]{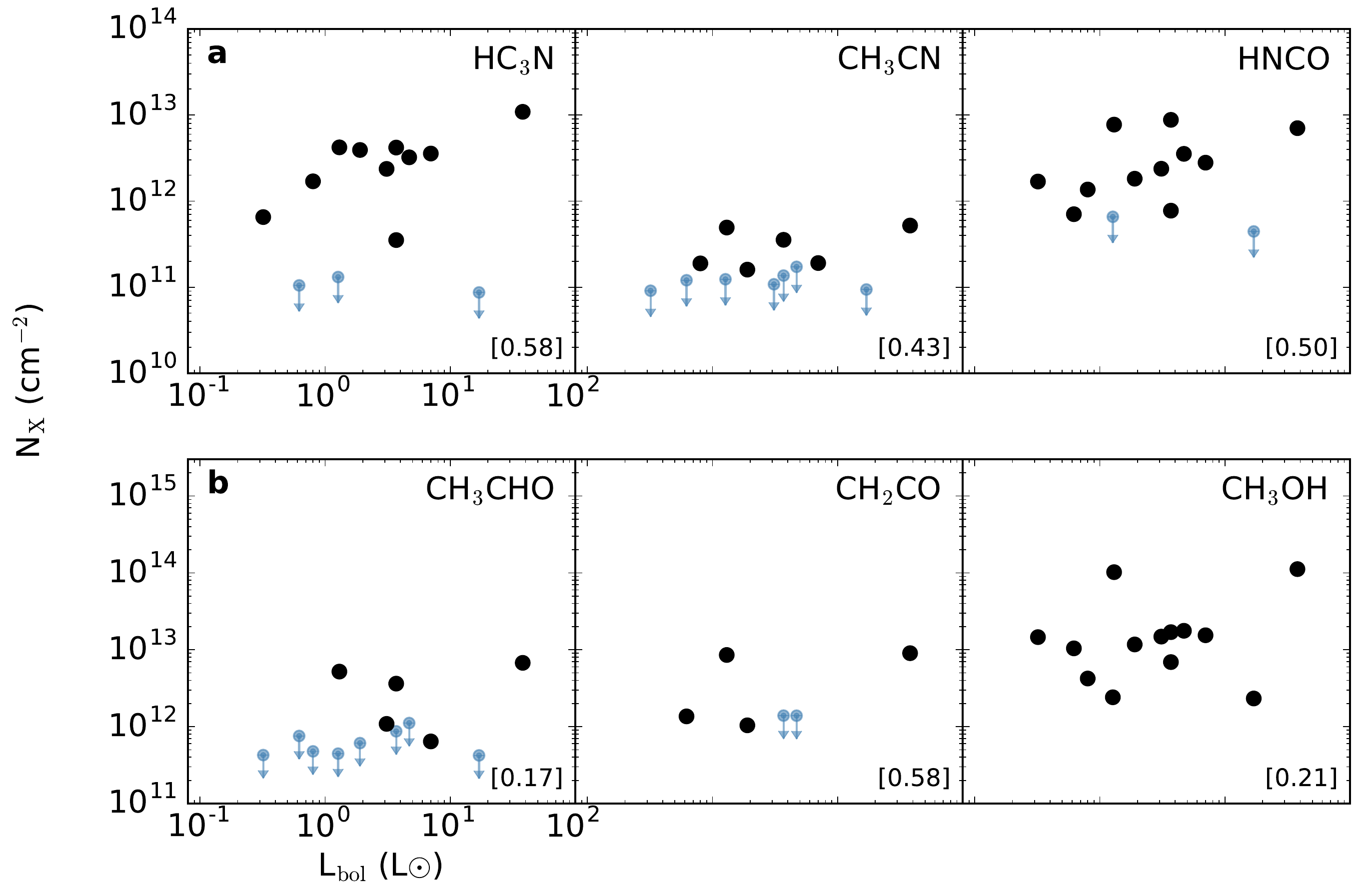}
	\caption{Column densities plotted against bolometric luminosity.  Upper limits due to molecule non-detections.  Pearson correlation coefficients are shown in brackets in the lower right corner.}
	\label{Lbol}
\end{figure} 
Correlations with bolometric luminosity are also positive, shown in Figure \ref{Lbol}.  In contrast to the envelope mass correlations, no single molecule stands out as particularly well-correlated with L$_{\mathrm{bol}}$.  This suggests that the weak correlations are not driven by chemistry, but rather excess excitation and ice desorption (affecting all molecules) around hotter protostars.  The most obvious outlier in Figures \ref{Menv} and \ref{Lbol} is L1448 IRS1, towards which no COMs are detected despite being one of the most massive and luminous of our sample.  Massive sources are more likely to be young sources \citep[e.g.][]{Hatchell2007}, so its unusual characteristics could be related to an early evolutionary stage.  

\begin{figure} 
	\centering
	\includegraphics[width=\linewidth]{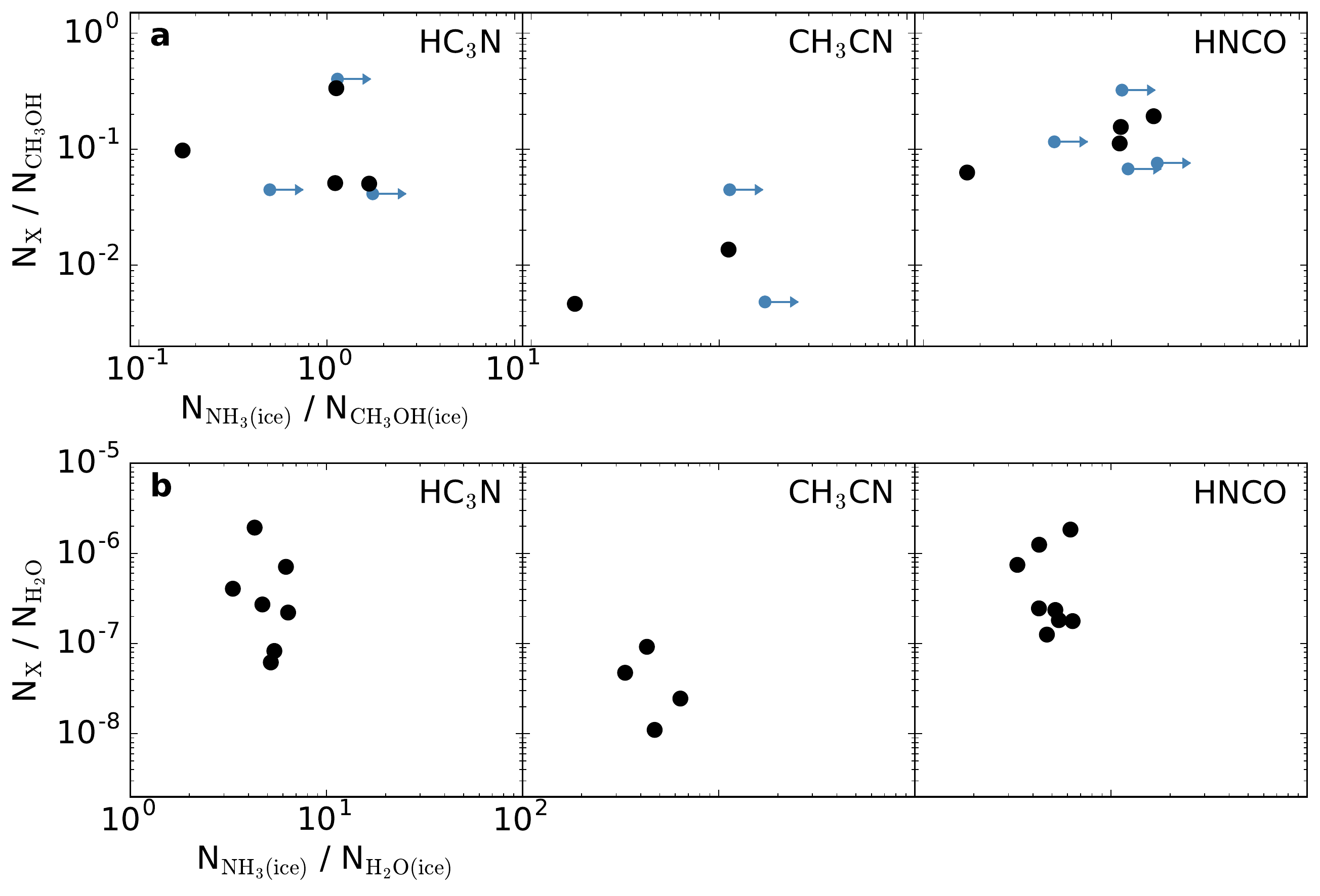}
	\caption{(a): Abundances with respect to CH$_3$OH for the nitrogen-bearing species plotted against the observed NH$_3$/CH$_3$OH ice ratio.  Lower limits due to non-detections of CH$_3$OH ice are shown in blue. (b): Abundances with respect to H$_2$O for the nitrogen-bearing species plotted against the observed NH$_3$/H$_2$O ice ratio.}
	\label{ices}
\end{figure} 

All sources in the sample have been detected in water ice and in at least one of the minor ice constituents NH$_3$, CH$_3$OH, and CH$_4$ (see Table 1).  In the context of COM formation, it is especially interesting to test whether the relative abundances of O- and N-bearing COMs are related to the relative ice abundances of CH$_3$OH (the theoretical starting point of O-rich COM chemistry) and NH$_3$ (the main observed carrier of nitrogen in interstellar ices \citep{Oberg2011}; N$_2$ is also a possible reservoir of nitrogen in ices but cannot be observed directly). The number of sources with both CH$_3$OH and NH$_3$ detections is unfortunately small: only 10 sources have both detections, or one detection and one low upper limit, for CH$_3$OH and NH$_3$ ice columns.  Of this, only a subset are rich in COMs, which limits the conclusiveness of any correlation studies. Figure \ref{ices}a shows the abundances with respect to CH$_3$OH gas for the nitrogen-bearing species, plotted against the NH$_3$/CH$_3$OH ice ratio towards the same sources.  HC$_3$N/CH$_3$OH gas-phase ratios are clearly not correlated with NH$_3$/CH$_3$OH ice ratios.  Otherwise, there are no conclusive relationships between the abundance of NH$_3$ ice and nitrogen-bearing molecules in this sample.  At first glace, CH$_3$CN/CH$_3$OH and HNCO/CH$_3$OH appear correlated with NH$_3$/CH$_3$OH ice ratios; however, the presence of lower limit outliers, one for CH$_3$CN and two for HNCO, suggests that the complex cyanide chemistry may be only weakly dependent on NH$_3$ ice abundances or altogether independent.  Further ice observations may reveal these lower limits to be anomalous, or may reveal relationships between other nitrogen-bearing molecules and NH$_3$ ice.  Figure \ref{ices}b shows the abundances with respect to H$_2$O instead of CH$_3$OH for NH$_3$ ice verses gas-phase nitrogen-bearing species.  Again there is no obvious relationship, indicating that the lack of dependency of cyanide chemistry on NH$_3$ is not an artifact of the normalizing species.

\section{Chemical Model}
\label{modeldesc}
\subsection{Model description}
In order to compare our observational results for LYSO chemical abundances with current theoretical predictions, we have used the three-phase chemical kinetics code MAGICKAL \citep[]{Garrod2013} to simulate LYSO chemistry. The model version and the chemical network employed are those presented by \citet{Garrod2017}; however, for the molecules of interest here, the differences between the work of \citet{Garrod2013} and more recent implementions are very minor. The model considers both the gas-phase and grain-surface/ice-mantle formation and destruction of molecules. Complex organic molecules are largely assumed to have only dust grain-related formation mechanisms, in the absence of other information.

We run the model multiple times using a selection of physical conditions that are representative of values at a range of radial distances from the core center. We use results from this grid of chemical models to produce a composite, spatially-dependent picture of the chemistry in the LYSO.

We begin with a spherically symmetric physical model of a generic LYSO, with a power-law temperature profile adapted from \citet{Chandler1999}:
\begin{equation}
T(r) = 60\Bigg{(}\frac{r}{2 \rm{x} 10^{15} m}\Bigg{)}^{-q}\Bigg{(}\frac{L_{bol}}{10^5 L_\odot}\Bigg{)}
^{q/2}  \rm{K},
\label{tprof}
\end{equation}
where we assume $q$ = 2/5 and $L_{bol}$ is either 1 or 10 L$_\odot$.  For the density profile, we assume the power law:
\begin{equation}
n_H(r) = n_{1000AU}\Bigg{(}\frac{r}{1000 AU}\Bigg{)}^{-\alpha}.
\label{pprof}
\end{equation}
Based on the median values determined from radiative transfer modeling of low-mass protostars in \citet{Jorgensen2002}, we assume $n_{1000AU}$ = 10$^{6}$ cm$^{-3}$ and $\alpha$ = 1.5.  We also assume an inner radius where T = 250K, again following \citet{Jorgensen2002}, and an outer radius where T = 10K. In the $L_{bol}$ = 10 L$_\odot$ case, a temperature of 250K is achieved at a radius of 3.8 AU, with a corresponding density of $4.27 \times 10^{9}$ cm$^{-3}$ at that position. In the $L_{bol}$ = 1 L$_\odot$ case, the peak temperature of 250K corresponds to a radius of 1.2 AU, at which a density of $2.89 \times 10^{10}$ cm$^{-3}$ is achieved. The latter value is taken as the maximum density for which chemical-model data are required (in either luminosity case). 

To populate the radial density profile with chemical data, we determine a total of 51 densities for which models are to be run, ranging logarithmically from the maximum of $2.89 \times 10^{10}$ cm$^{-3}$ down to $3.45 \times 10^{4}$ cm$^{-3}$. For each of these densities, we run a dedicated hot-core type model, using a two-stage approach \citep[following ][]{Garrod2013}: stage one consists of an isothermal collapse from density 3000 cm$^{-3}$ up to the final density chosen for that run; in stage two, this density is fixed and the gas and dust temperatures gradually rise to a maximum of 400K, using the ``intermediate'' warm-up timescale (which goes from 8 -- 200K in $2 \times 10^{5}$ yr, producing a total warm-up timescale from 8 -- 400K of $2.85 \times 10^{5}$ yr).

For either the $L_{bol}$ = 1 L$_\odot$ or 10 L$_\odot$ case, models with densities that fall within the maximum and miminum values are chosen and placed at the appropriate radius. Then, from within each model, the abundances of all simulated molecules are extracted according to the instantaneous temperature achieved during warm-up, as determined by the temperature profile at that specific radius. To obtain data at the precise temperatures required, the model output data are interpolated between output temperature values; the temperature resolution obtained in the models is always less than 1\% of the absolute value.

In this way, each model placed at a specific position in the density profile is fixed in time according to the local temperature. To account for uncertainties in the temperature profile, we also extract chemical profiles corresponding to local temperature values that are 10\% lower and 10\% higher temperatures than those given by the fiducial temperature profile in equation \ref{tprof}.  

\begin{figure}
	\centering
	\includegraphics[width = 0.7\linewidth]{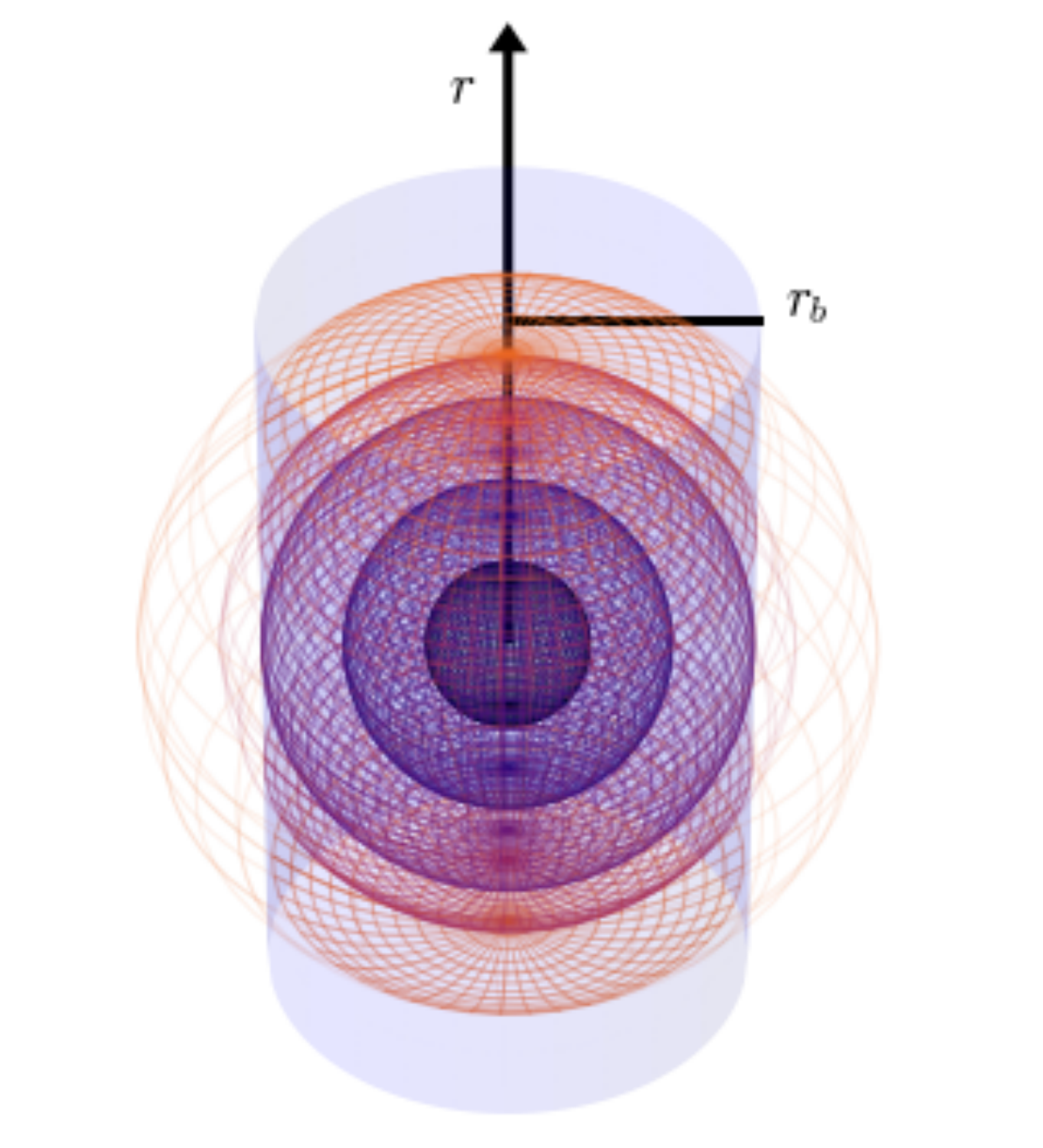}
	\caption{Schematic of emitting surfaces contained within a cylindrical beam for different radial shells.}
	\label{beamsim}
\end{figure}

The resulting radial profiles for all molecules are then used to calculate the line-of-sight column densities that would be observed with a single-dish telescope. To do so, we convolve the molecular abundance $\frac{n_X}{n_H}(r)$, hydrogen density $n_H(r)$, and emitting area $A(r)$ at each radius.  We then integrate this from our minimum to maximum radii to yield the total number of molecules of a given species within the beam:
\begin{equation}
n_X = \int_{r_{min}}^{r_{max}}A(r)n_H(r)\frac{n_X}{n_H}(r).
\label{integr_CD}
\end{equation}  
The column density is then simply:
\begin{equation}
N_X = \frac{n_X}{\pi r_b^2},
\end{equation}
where $r_b$ is the beam radius.  We assume a 10" beam radius based on the size of the observing beam (Section \ref{sec_obs}), and a distance to the model protostar of 200pc.  In order to account for the cylindrical line of sight, we express $A(r)$ as:
\begin{equation}
A(r) = 
\begin{cases}
4\pi r^2, & \text{if } r \le r_{b} \\
2\pi [r_{b}^2 + \big{(}r - \sqrt{r^2 - r_{b}^2}\big{)}^2 ], & \text{if } r > r_{b}
\end{cases}
\label{SA}
\end{equation}
In other words, while the radius is smaller than the beam radius, the emitting surface corresponds to the surface area of a sphere at that radius.  Beyond the beam radius, the emitting surface consists of spherical caps in front of and behind a sphere of radius = $r_b$.  These caps have a constant fixed base radius $r_b$ and a height that varies depending on the radius.  Figure \ref{beamsim} illustrates how this treatment accounts for material in the line of sight while excluding material outside the beam. 

Our use of a single warm-up timescale keeps the model grid to a manageable size.  The choice to use the ``intermediate'' timescale is based on its ability to best reproduce the abundances of hot-core type molecules at the high temperatures of such sources.  In the case of the low-mass protostars that concern this work, the ``fast'' timescale could also be an appropriate choice; however, the differences between the ``fast'' and ``intermediate'' results are mainly quantitative rather than qualitative (in contrast with the ``slow'' warm-up timescale models).  Differences between peak abundances calculated by these models are likely to be on the order of a factor of a few.  Such variations are acceptable in the context of our generic model, but the modeling of specific sources in the future may be tailored to more specific warm-up timescales, using an explicitly determined infall speed.

\subsection{Model results}
\begin{figure*}
	\centering
	\includegraphics[width = 0.8\linewidth]{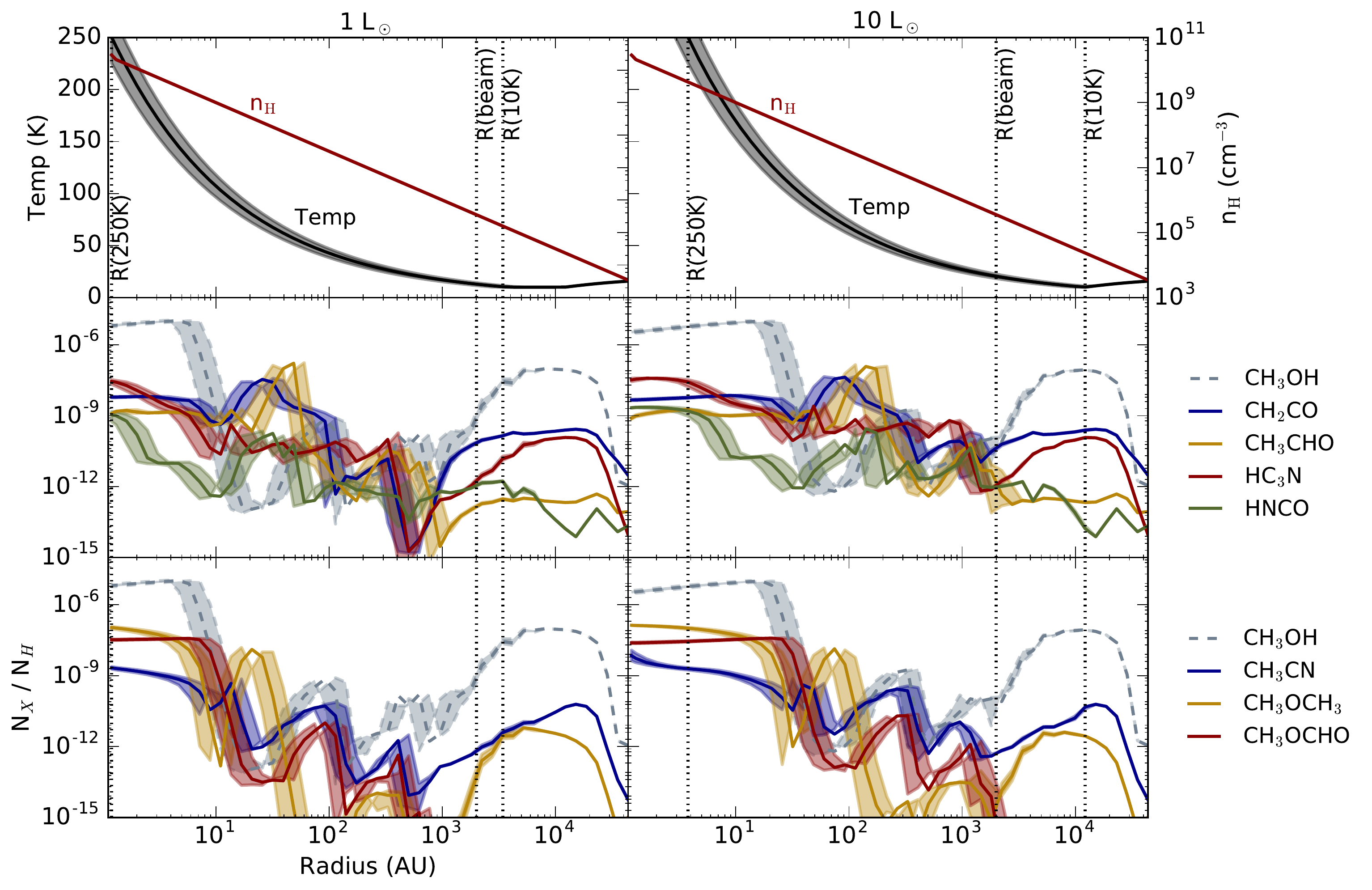}
	\caption{Chemical models for 1L$_\odot$ and 10L$_\odot$ LYSO simulations.  In all panels, dotted vertical lines mark the inner radius at 250K, the beam radius of 10" for an object at 200pc, and the outer radius for a 10K cutoff temperature. The shaded regions around each line represent results from temperature profiles 10\% higher and 10\% lower than the fiducial run.  Top panels: temperature and density profiles (equations \ref{tprof} and \ref{pprof}).  Middle panels: fractional abundances of the typically "cold" emitters CH$_2$CO, CH$_3$CHO, HC$_3$N, and HNCO.  Bottom panels: abundances of the typically "warm" emitters CH$_3$CN, CH$_3$OCH$_3$, and CH$_3$OCHO. CH$_3$OH is shown for comparison in both the middle and bottom panels.}
	\label{modelprof}
\end{figure*}

Figure \ref{modelprof} shows the 1D radial profiles of COM fractional abundances for both the 1L$_\odot$ and 10L$_\odot$ protostar simulations. The variations in temperature by $\pm 10$\% do not produce significant changes in the peak abundances; rather, they alter the radial onset of peak values, thereby changing the calculated column densities.  Generally, trends in the radial abundances are consistent with our expectations based on observations: CH$_3$OCH$_3$ and CH$_3$OCHO, which are thought to emit from warm inner regions of the protostar, drop off in abundance moving to larger radii, while the "cold" emitters CH$_2$CO, CH$_3$CHO, HC$_3$N, and HNCO maintain a high abundance at all radii.  Interestingly, CH$_3$CN has a profile more similar to the cold emitters despite its observationally derived warmer rotational temperature.  We discuss the chemistry of these models further in Section \ref{models}.

\section{Discussion}
\label{sec_disc}
\subsection{COM formation chemistry}
\label{chem}
Correlations among COMs and between COMs and source properties provide constraints on their chemistry.  In this subsection, we discuss the formation chemistry of individual molecules based on their rotational temperatures and observed correlations with CH$_3$OH, protostellar properties, and ice abundances (Figures \ref{fig_ch3ohcorr}-\ref{ices}).

\textit{CH$_2$CO:} CH$_2$CO formation has been demonstrated experimentally by ice processing, with proposed formation mechanisms of either C$_2$O hydrogenation \citep{Maity2014} or vinyl alcohol formation and decomposition \citep{Hudson2013}.  However, theory indicates that H addition to C$_2$O is the dominant pathway \citep{Garrod2008}.  The strong correlation of CH$_2$CO with CH$_3$OH along with its low rotational temperature suggest that ketene forms via atom addition reactions in simple ices (i.e. CO + C $\rightarrow$ C$_2$O, followed by hydrogenation to CH$_2$CO) similar to CH$_3$OH formation.

\textit{CH$_3$CHO:} CH$_3$CHO also correlates strongly with CH$_3$OH.  CH$_3$CHO has a very low excitation temperature (8K) and is therefore likely a zeroth or first-generation COM.  As mentioned in Section \ref{sec_PCC}, CH$_3$CHO does not correlate well with M$_{\mathrm{env}}$; this is in contrast to the gas-phase chemistry product HC$_3$N.  CH$_3$CHO is therefore not likely to be a gas-phase product.  Both experiments \citep{Bennett2005,Oberg2009a} and modeling \citep{Garrod2008} demonstrate efficient formation of acetaldehyde from radical products of CH$_3$OH, namely CH$_3$ and HCO.  CH$_3$OH processing is not the only pathway to CH$_3$CHO (CH$_3$ can also be derived from the dissociation of CH$_4$, and the HCO radical can form from CO hydrogenation or dissociation of H$_2$CO), but the tight correlation suggests that CH$_3$OH is an important precursor to CH$_3$CHO

\textit{CH$_3$OCH$_3$ and CH$_3$OCHO:}  Methyl formate and dimethyl ether have fairly warm rotational temperatures in our survey, with averages of 17K and 16K, respectively.  This is much warmer than CH$_3$CHO, suggesting that these molecules originate closer to the central protostar.  Both molecules are thought to form from recombinations of radical products of methanol, and our data is consistent with this formation scenario: CH$_3$OCH$_3$ and CH$_3$OCHO are the most abundant towards sources with high CH$_3$OH column densities.  However, with only two detections this cannot be ascertained with any confidence.  Firm detections in sources with lower column densities will help to elucidate this relationship.  We note that this will likely require ALMA observations, as the observations presented here are already quite deep.  Spatially resolved observations towards the sources with detections would also be helpful in constraining the formation chemistry.

\textit{HNCO:}  The chemical origin of HNCO is still not particularly well understood.  Recent experiments have shown that HNCO can be produced in ices from CO reacting with NH radicals before N is fully hydrogenated to NH$_3$ \citep{Fedoseev2015}, or from simultaneous UV irradiation and hydrogenation of NO in CO, H$_2$CO, and CH$_3$OH ices \citep{Fedoseev2016}.  Modeling by \citet{Garrod2008} shows that HNCO formation is efficient only by the gas-phase destruction of more complex grain-surface products, urea in particular.  However, the recent model of \citet{Belloche2017} suggests reaction between NH and CO may be efficient (see Section \ref{models}).  Several aspects of our HNCO observations support a scenario in which HNCO mainly forms through atom addition ice chemistry in the cold envelope.  First, the ubiquity of HNCO in our sources favors an early formation (i.e. zeroth-generation COM chemistry): its detection towards sources with no other detections precludes a formation chemistry dependent on high gas-phase abundances of large complex molecules (i.e. second-generation formation).  Moreover, HNCO correlates relatively well with envelope mass, indicative of formation in the envelope.  Finally, HNCO column densities are fairly well correlated with CH$_3$OH for sources with low column densities of CH$_3$OH, but the two sources with much higher CH$_3$OH column densities (B1-a and SVS 4-5) do not display a comparable increase in HNCO column density.  This is consistent with a scenario of co-formation of HNCO and CH$_3$OH in the cold envelope, while HNCO destruction in the core region could explain the saturation behavior.  

\textit{CH$_3$CN:}  CH$_3$CN has the highest rotational temperature of all the COMs observed, with an average of 27K; together with the lack of correlation between CH$_3$CN and M$_{\mathrm{env}}$, this suggests a more centrally concentrated emission origin and an efficient "lukewarm" pathway to CH$_3$CN.  In hot cores, CH$_3$CN is thought to form mainly on grain surfaces through recombination of CH$_3$ + CN, followed by sublimation above $\sim$90K \citep{Garrod2008}.  A gas-phase channel, involving HCN + CH$_3^+$ $\rightarrow$ CH$_3$CNH$^+$ followed by CH$_3$CNH$^+$ + e$^-$ $\rightarrow$ CH$_3$CN + H, may become important at lower temperatures.  In our sample, CH$_3$CN correlates well with CH$_3$OH, indicating that these molecules have similar underlying chemistries and therefore that ice chemistry may be important in CH$_3$CN formation.  However, as discussed in Section \ref{models}, the chemical model relies on gas-phase chemistry to efficiently form CH$_3$CN at these temperatures.  Thus, there is some tension between the model and observations in inferring the dominant CH$_3$CN formation channel.  Finally, CH$_3$CN has no obvious correlation with the NH$_3$ ice column, indicating that cyanide production in the ice is not linked to NH$_3$.

\textit{HC$_3$N:}  HC$_3$N has the most scatter of any molecule in its correlation with CH$_3$OH.  This is unsurprising given that cyanopolyyne observations can be well-reproduced by purely gas-phase models \citep{Herbst2009}; any correlation with CH$_3$OH is likely related to the scale of the object and not a reliance on CH$_3$OH chemistry.  This is also consistent with the weak correlation of HC$_3$N with all other COMs, discussed in Section \ref{sec_PCC}.  The tight correlation between envelope mass and HC$_3$N abundance and the obvious lack of correlation with NH$_3$ ice further support this.  These observations support the general notion that hydrocarbon species are mainly gas-phase products, and do not require grain-surface chemistry to explain their abundances.  Interestingly, \citet{Graninger2016} found a positive correlation between the carbon chain C$_4$H and CH$_3$OH in these LYSOs; this suggests that nitrogen-bearing carbon chains have distinct gas-phase formation chemistries from unsubstituted carbon chains. 

\subsection{Modeled vs. observed chemistry}
\label{models}
\begin{figure}
	\centering
	\includegraphics[width = \linewidth]{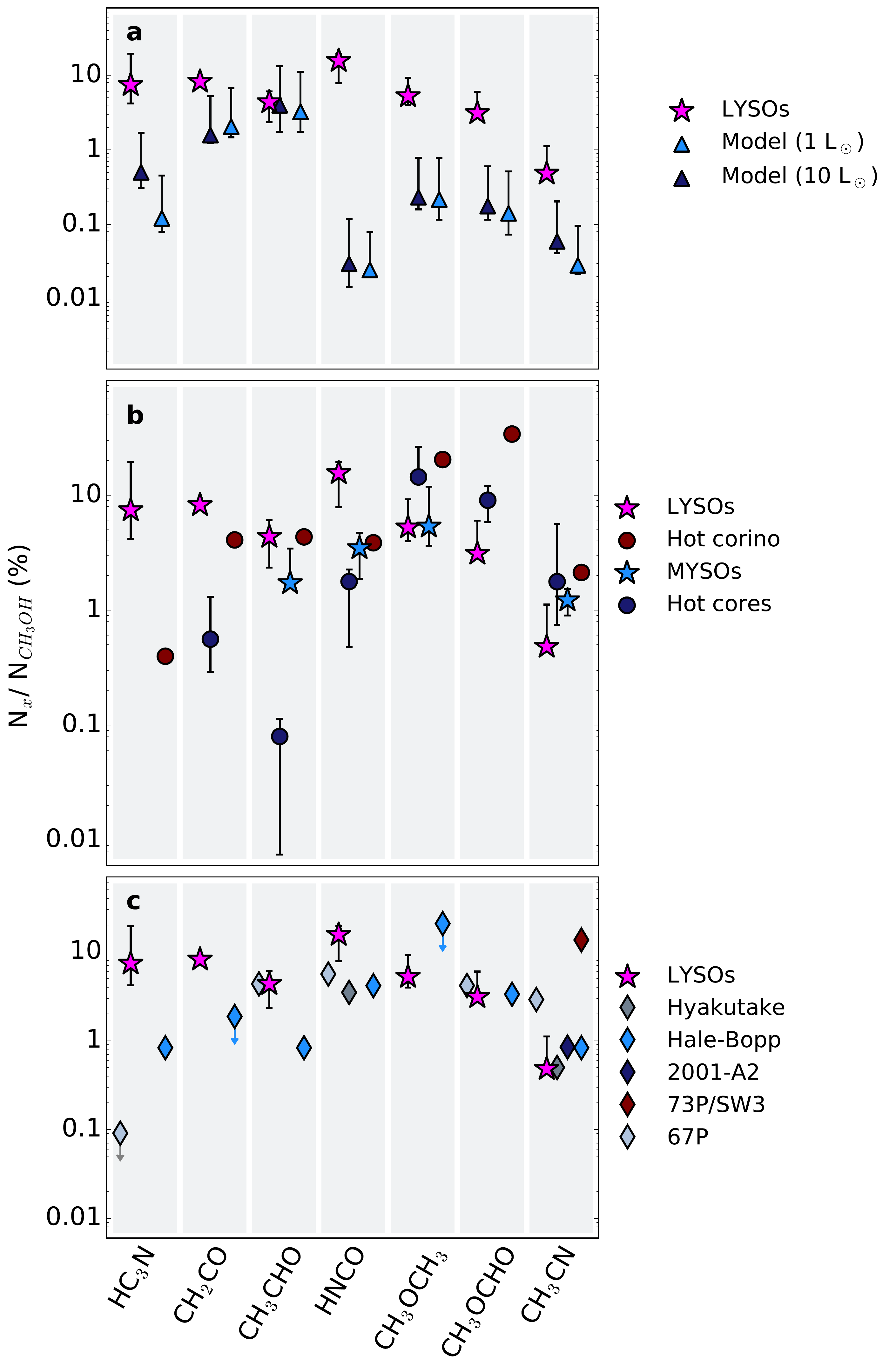}
	\caption{COM abundances with respect to CH$_3$OH across different types of objects. For all panels, LYSOs are shown as pink stars.  Scatterpoints represent median values and error bars span the first and third quartile.  a: Comparison with abundances from the protostellar warm-up model described in Section \ref{modeldesc} for 1 L$_\odot$ (light blue) and 10 L$_\odot$ (dark blue) profiles.  Error bars span the results for temperature profiles 10\% higher and 10\% lower than the fiducial run.  b: Comparison of different star formation environments.  Hot corinos \citep{VanDishoeck1995,Cazaux2003} are shown in dark pink; MYSOs \citep{Fayolle2015} in light blue; and hot cores \citep{Bisschop2007a} in dark blue.  c: Comparison with observations of solar system comets \citep{Crovisier2004,Mumma2011,Goesmann2015}.}
	\label{fig_compare}
\end{figure}

Figure \ref{fig_compare}a compares beam-averaged COM abundances with respect to CH$_3$OH for our observations and modeling results (Section \ref{modeldesc}).  For the observations, median values are shown as pink stars, with error bars spanning the first and third quartile.  For the model, we show results from both 1L$_\odot$ (light blue) and 10L$_\odot$ (dark blue) temperature profiles.  Triangles represents results from the fiducial temperature profiles, and error bars show the results for temperature profiles 10\% higher and 10\% lower. 

Apart from HNCO, the model abundances are within an order of magnitude of the observations.  Agreement within an order of magnitude is very reasonable given modeling uncertainties in the reaction rates as well as observational uncertainties in the different excitation properties of the molecules being compared.  In particular, the systematic underestimation of most molecules by the model may be due in part to an excess of CH$_3$OH in the model at large radii, caused by overactive chemical desorption of methanol at cold temperatures.  The underproduction of HNCO by several orders of magnitude is likely a result of an over-estimated barrier for the grain surface process of NH + CO $\rightarrow$ HNCO, which impedes efficient grain-surface formation of HNCO in the model.  While there are no strong constraints on this barrier from laboratory work, recent modeling work by \citet{Belloche2017} tested several values ranging from 1000 -- 2500K, the latter value being the one used in our present study.  They found that the barrier is required to be no greater than $\sim$1500K in order to produce sufficient HNCO to explain observed abundances of CH$_3$NCO. The value of around 2500K used in the present model was adopted by \citet{Garrod2008}, when this reaction was first introduced into the network, and was based on the barrier found for the analogous reaction H + CO $\rightarrow$ HCO. Reducing the current value of 2500K to the 1500K recommended by Belloche et al. may therefore improve agreement with our observational data in future model iterations.

\begin{figure}
	\centering
	\includegraphics[width = \linewidth]{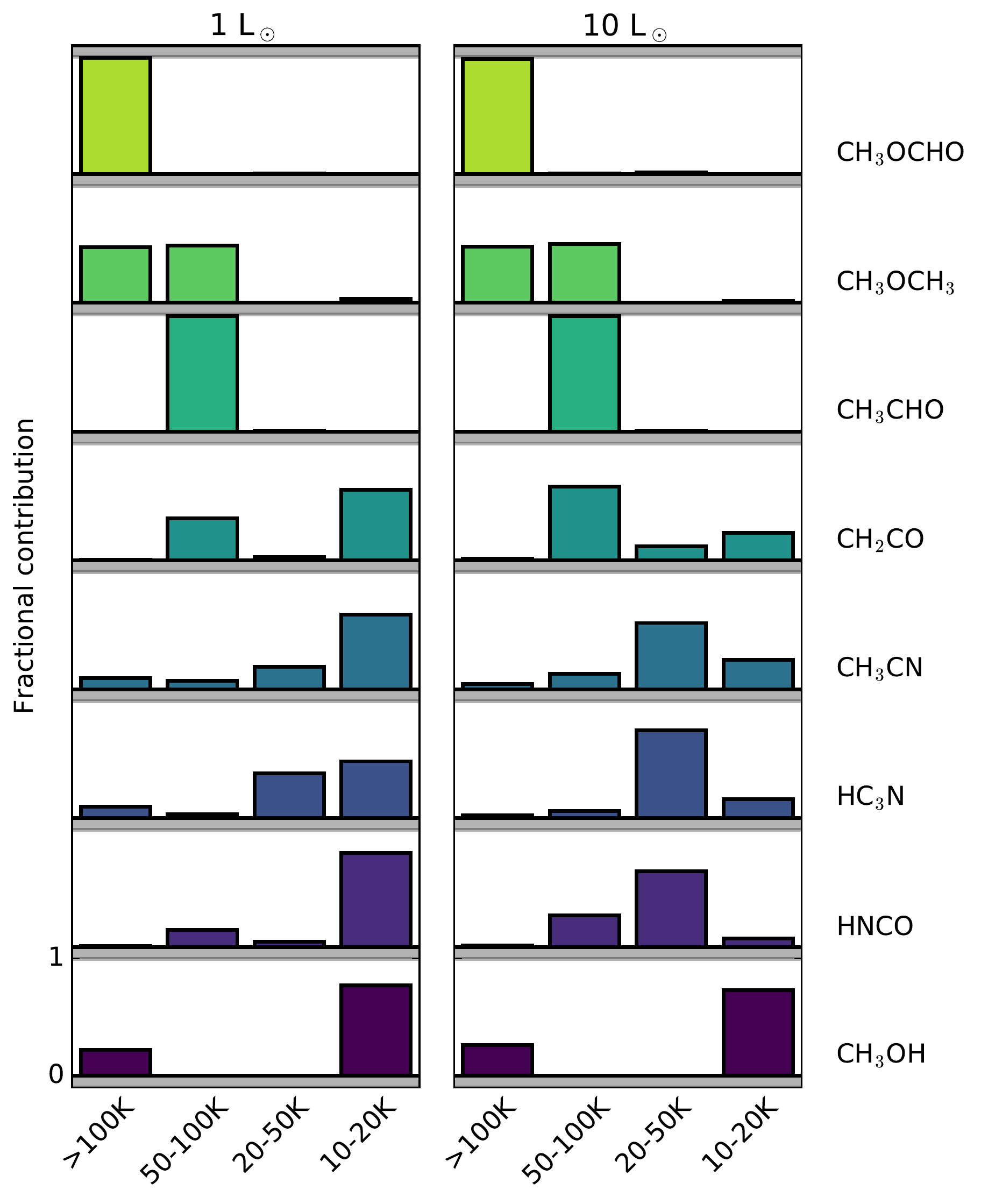}
	\caption{Model results showing the fraction of each molecule's total column density that comes from different temperature ranges.  For each molecule, the y scale is the same as is indicated for CH$_3$OH.}
	\label{mol_bins}
\end{figure}

Figure \ref{mol_bins} shows the fraction of each molecule's total column density coming from different temperature ranges in the model, which can be compared with measured rotational temperatures.  We note that if the observed line transitions are not in LTE then the measured rotational temperatures will underestimate the kinetic temperature.  To assess how much this impacts our observations, we run RADEX models for molecules with both collisional constants and rotational temperatures \citep[HC$_3$N, CH$_3$CN, and CH$_3$OH;][]{vanderTak2007} using the measured rotational temperatures and column densities as inputs and assuming the density and temperature profiles in Section 4.1.  We find that CH$_3$CN and HC$_3$N observations are well reproduced by the simulated RADEX emission while CH$_3$OH is not.  Therefore, for some molecules (especially oxygen-bearing species), the measured rotational temperatures and column densities may be affected by non-LTE effects.
	
We also estimate the temperature above which HC$_3$N, CH$_3$CN, and CH$_3$OH should be in LTE.  This is done using the collisional constants for these molecules \citep{Schoier2005}, again assuming the model density and temperature profiles.  We find that most lines transition to non-LTE conditions at densities corresponding to 10-25K.  Most molecules have observed rotational temperatures around or below this range; this suggests that at least some component of our observed emission is sub-thermal, consistent with the RADEX results.  Therefore, in comparing with the model emission temperatures, we emphasize that the measured rotational temperatures should be regarded as broadly corresponding to inner, outer, and intermediate emission origins, and likely do not correspond exactly to the actual kinetic temperature.

CH$_3$OCHO and CH$_3$OCH$_3$ are both found at warm to hot temperatures ($>$50K) in the model, compared with rotational temperatures of 16K and 17K, respectively.  However, these molecules are detected towards only 2 sources, and only in a few detected lines, so it is difficult to assess how serious these constraints are.  They seem to rule out, however, that the emission is coming exclusively from the hot material as predicted by the model.  HC$_3$N, HNCO, CH$_2$CO, and CH$_3$CN are all dominated by temperatures colder than 50K in the model.  The available rotational temperatures (14K for HC$_3$N and 27K for CH$_3$CN) are fully consistent with this.  Thus, the model successfully reproduces the envelope-dominated, low-temperature chemistry seen in these LYSOs, without the need for hot-core chemistry.  This is true even for CH$_3$CN, which is typically considered a hot core molecule.  The gas-phase pathway described in Section \ref{chem} is mainly responsible for this lukewarm formation in the model, and is driven by an enhancement of gas-phase HCN.  This in turn is due to either sublimation of HCN from grains in warmer regimes ($\sim$40K), or purely gas-phase HCN formation in cooler regimes.  The observed rotational temperature of 27~K is consistent with some combination of low and intermediate temperature production of CH$_3$CN.

CH$_3$CHO is found entirely at warm temperatures (50-100K) in the model due to desorption of CH$_3$CHO formed on grain surfaces; this is much higher than its observed 8K rotational temperature.  Unlike methyl formate and dimethyl ether, acetaldehyde has very well-populated rotational diagrams and therefore the rotational temperature is more secure.  It seems, then, that the observed CH$_3$CHO is being produced through some cold formation pathway other than the radical recombination chemistry included in the model.  

The majority of CH$_3$OH in the model is found at cold temperatures (10-20K).  Cold CH$_3$OH in the model is likely somewhat over-abundant due to an overactive chemical desorption, contributing to the systematic under-estimation of other COM abundances seen in Figure \ref{fig_compare}.  Even so, observational rotational temperatures for CH$_3$OH in these sources are $\sim$5-6K \citep{Graninger2016}, consistent with a mainly cold origin.  This suggests that while the magnitude of cold CH$_3$OH in the model may be too high, the trend of dominantly cold emission is appropriate.

The high efficiency of cold chemical desorption of methanol is a result of the rapid abstraction of hydrogen from methanol by H atoms, followed by re-hydrogenation by another H atom. Chemical desorption occurs in 0.24 \% of re-hydrogenation cases, as determined by the method of \citet{Garrod2007}, using an efficiency factor $a=0.01$. The abstraction and re-hydrogenation of methanol on the grains in this model is sufficiently fast to produce an excess of gas-phase methanol, while the total remaining on the grains is affected only to a relatively small degree. The CO-H$_2$CO-CH$_3$OH system on grains is complex, with reactions that allow inter-conversion in both directions. The barriers and branching ratios for many of these processes are poorly-defined. The return of grain surface-formed methanol to the gas phase is therefore a function of both its formation/re-formation on grains and the efficiency of desorption, which is also poorly constrained.

In addition to gas-phase abundances, the model also provides ice abundances.  In practice, observations of ices are made in absorbance and therefore trace a pencil beam of protostellar material rather than the entire envelope.  To compare with observed ice column densities, therefore, the number density of each ice species in the model is simply integrated out to the maximum radius.  The model produces fairly consistent values with the measured ice columns in Table 1, with H$_2$O column densities on the order 10$^{19}$cm$^{-2}$ compared with observed values of 10$^{18}$-10$^{19}$cm$^{-2}$.  CH$_3$OH model ice abundances with respect to water are around 7\%, which is consistent within a factor of a few of measured values, while NH$_3$ model abundances of 19\% are somewhat high compared to observations.
	
In the model, the number densities of ice species peak just outside of the ice sublimation line around 130K, corresponding to 6AU in the 1L$_\odot$ case and 21AU in the 10L$_\odot$ case.  This is a much smaller spatial scale than our gas-phase observations, and therefore ice absorption measurements are likely not probing the same radius as the gas-phase observations.  Nonetheless, observed gas and ice abundances should still be related for molecules that form primarily in the ice phase, assuming that ice compositions do not evolve substantially between the outer and inner envelope.  Indeed, in the model the ice abundance profiles are essentially flat from the 10K radius in to the sublimation line.  Ices observed in absorption close to the sublimation line should therefore preserve a similar composition to ices in the outer envelope, from which many of the observed gas-phase COMs likely originate.

\subsection{Massive vs. low-mass protostellar chemistry}
\label{star}
We next compare observations of COMs towards different types of protostars.  We consider our sample of low-mass protostars without clear evidence for hot corinos (subsequently termed LYSOs), the hot corino source IRAS 16293-2422, and analogously high-mass protostars without and with hot cores (termed MYSOs and hot cores, respectively).  Abundance measurements for the hot corino are taken from \citet{VanDishoeck1995} and \citet{Cazaux2003}; the MYSO sample is from \citet{Fayolle2015}; and the hot-core sample is from \citet{Bisschop2007a}.  Survival analysis was performed on the MYSO and hot-core samples in order to derive median values accounting for upper limits in these surveys.  

We adopt beam-averaged abundances with respect to CH$_3$OH under the assumption that the COMs and CH$_3$OH have the same distributions \citep{Herbst2009,Oberg2011a}.  This may not be an appropriate assumption, particularly for centrally concentrated warm emitters, but a lack of consistent structural information for the different sources prevents a more thorough treatment of beam dilution.  There are some uncertainties then in the abundances with respect to CH$_3$OH due to potential differences in the emission origin of each COM.  

We now compare our LYSO sample with the hot corino IRAS 16293.  We note that the LYSO sources in our survey could potentially host hot corinos, however our observations are not sensitive to this emission.  The maximum observed transition energy of $\sim$40K as well as the exclusion of line wings from spectral fitting support that the molecules observed are emitting from the outer regions of the protostar.  Compared with IRAS 16293, the LYSOs are enhanced in the carbon chain HC$_3$N, comparable or slightly enhanced in the cold COMs CH$_2$CO, CH$_3$CHO, and HNCO, and underabundant in the hot COMS CH$_3$CN, CH$_3$OCH$_3$, and CH$_3$OCHO.  This suggests an evolutionary sequence that results in the appearance of a central brightly emitting hot corino without destroying the envelope COM chemistry.    We note that IRAS 16293 may not be representative of typical hot corino abundances, and a larger sample of these objects is required to make firm comparisons.  

The MYSOs in \citet{Fayolle2015} have large envelopes with ice detections and lack a central hot core, and should therefore represent a comparable evolutionary stage to our LYSO sample.  LYSOs have slightly higher abundances relative to MYSOs of the cold molecules CH$_3$CHO and HNCO, and comparable abundances for the warm molecules CH$_3$OCH$_3$ and CH$_3$CN.  Overall, the chemistries of these types of objects appear very similar, with slight enhancements of cold molecules in the LYSOs that could be attributed to the generally colder temperatures of low-mass protostars.

It has been previously noted that hot corinos of LYSOs have abundances comparable to or greater than hot cores \citep[e.g][]{Bottinelli2007,Herbst2009}.  While this is true for IRAS 16293, we note that from our unbiased sample of LYSOs this is clearly not typical for all low-mass objects.  Furthermore, generalizing COMs as a single category seems to be an overly reductive classification, as cold and hot molecules represent distinct chemistries, and their emission trends between environments may differ.

\subsection{Comparisons with solar system chemistry}
\label{comet}
Finally, we compare LYSO median abundances with observations of solar system comets \citep{Crovisier2004,Mumma2011,Goesmann2015}.  In Figure \ref{fig_compare}c, most cometary abundances appear consistent with LYSO abundances, with the exception of HC$_3$N and CH$_2$CO.  Since LYSOs represent regions of solar-type star formation, this similarity supports the notion that (1) the solar system is fairly typical in its chemistry, and other sun-like stars should have similar chemical inventories; and (2) chemistry early in the evolution of a star is propagated through later stages.  This suggests that some of the comet's molecular composition is inherited from the early stages of star formation.  

HC$_3$N and CH$_2$CO are each about an order of magnitude higher in LYSOs than in comets.  The formation mechanisms discussed in Section \ref{chem} suggest that it may be difficult to form these molecules at later evolutionary stages: HC$_3$N is a dominantly gas-phase product favored by the low-density, cold outer envelope, while CH$_2$CO is thought to form early via atom addition in simple ices.  Both of these are also unsaturated molecules that are susceptible to depletion via hydrogenation or photoprocessing.  This is in contrast to CH$_3$CHO and HNCO, which are also unsaturated but have formation channels involving radical recombinations that can continue at later stages of disk and planet formation.

\section{Conclusions}
Based on a survey of complex organic molecules towards 16 young low-mass protostars using the IRAM 30m telescope, we conclude the following:
\begin{enumerate}
\item The molecules CH$_2$CO, CH$_3$CHO, CH$_3$OCH$_3$, CH$_3$OCHO, HC$_3$N, and HNCO all have median column densities on the order of 10$^{12}$ cm$^{-2}$ and median abundances with respect to CH$_3$OH around five to ten percent.  CH$_3$CN is an order of magnitude lower in column density and abundance.
\item For this sample of LYSOs at similar evolutionary stages, COM column densities span at least an order of magnitude.  The distribution is reduced, especially for oxygen-bearing molecules, upon normalizaiton to CH$_3$OH.
\item Our findings are consistent with grain-surface chemistry as the dominant formation pathway to most COMs in this survey.  There is evidence that the complex organics either form from or co-form with CH$_3$OH.  While the nitrogen chemistry is regulated by additional factors, it appears still related to the CH$_3$OH chemistry.
\item A warm-up model for low-mass protostars produces COM abundances in fair agreement with observations: the relative abundances of most molecules follow the same trend as seen observationally, though with a systematic under-estimation due to an excess of CH$_3$OH in the model.  Most molecules are produced in the model by cold to lukewarm chemistry, consistent with observed rotational temperatures.  CH$_3$CHO is an exception, occurring at much higher temperatures in the model compared to observations.  Gas-phase chemistry is responsible for lukewarm CH$_3$CN production in the model, driven by high HCN abundances.
\item LYSO COM abundances are comparable to measurements in solar system comets, indicating that the sun is fairly typical among low-mass stars.
\end{enumerate}

The study is based on observations with the IRAM 30m Telescope. IRAM is supported by INSU/CNRS (France), MPG (Germany) and IGN (Spain).  J.B.B acknowledges funding from the National Science Foundation Graduate Research Fellowship under Grant DGE1144152. K.I.O. acknowledges funding from the Simons Collaboration on the Origins of Life (SCOL) investigator award.

\newpage
\section{Appendix: Detection spectra}
\begin{figure*}[h!]
	\centering
	\includegraphics[width=0.8\textwidth]{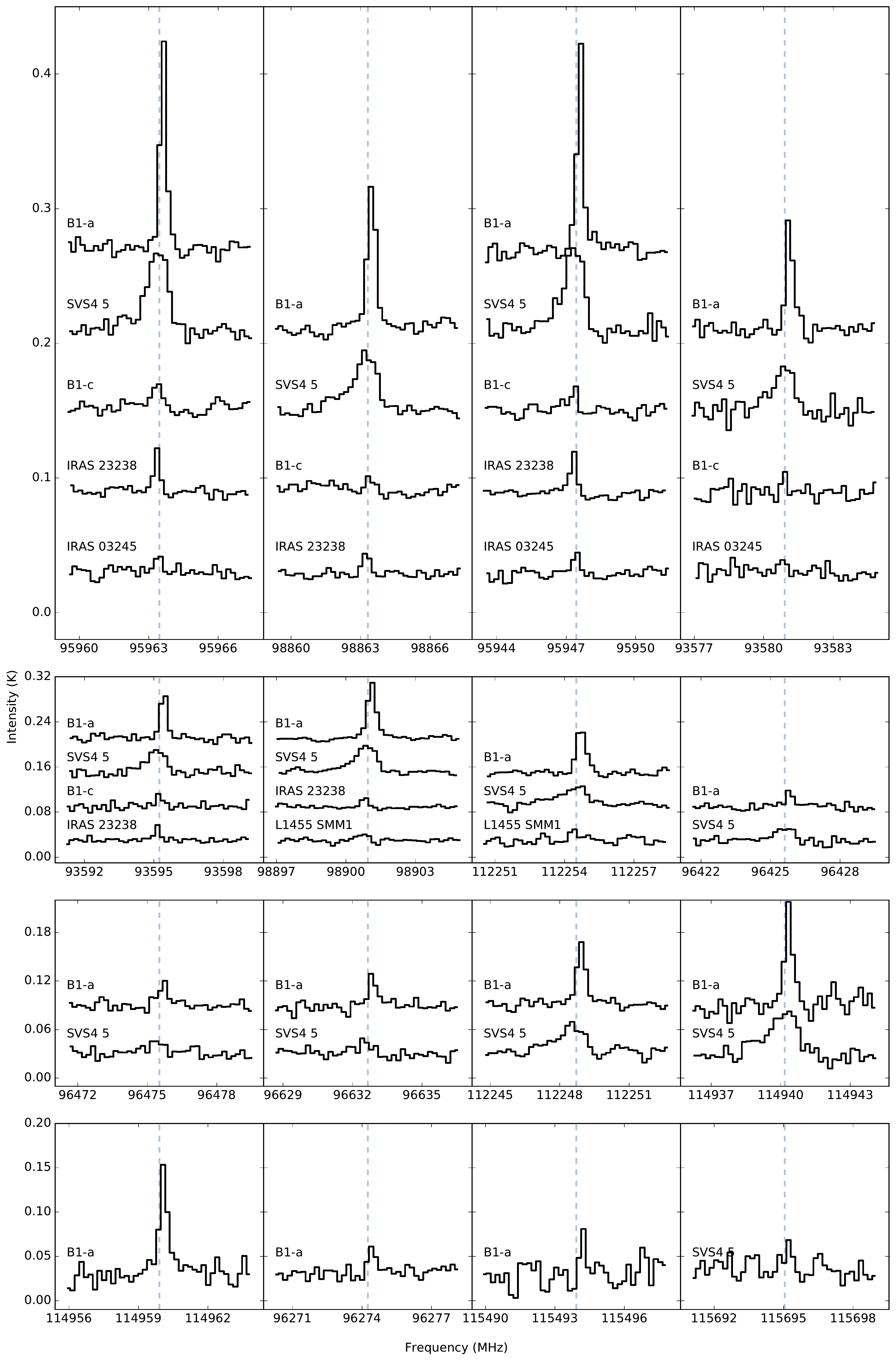}
	\caption{CH$_3$CHO detected transitions.  Grey dashed lines show line centers; spectra are offset for clarity.}
	\label{spec_ch3cho}
\end{figure*}
\begin{figure}[h!]
	\centering
	\includegraphics[width=0.64\linewidth]{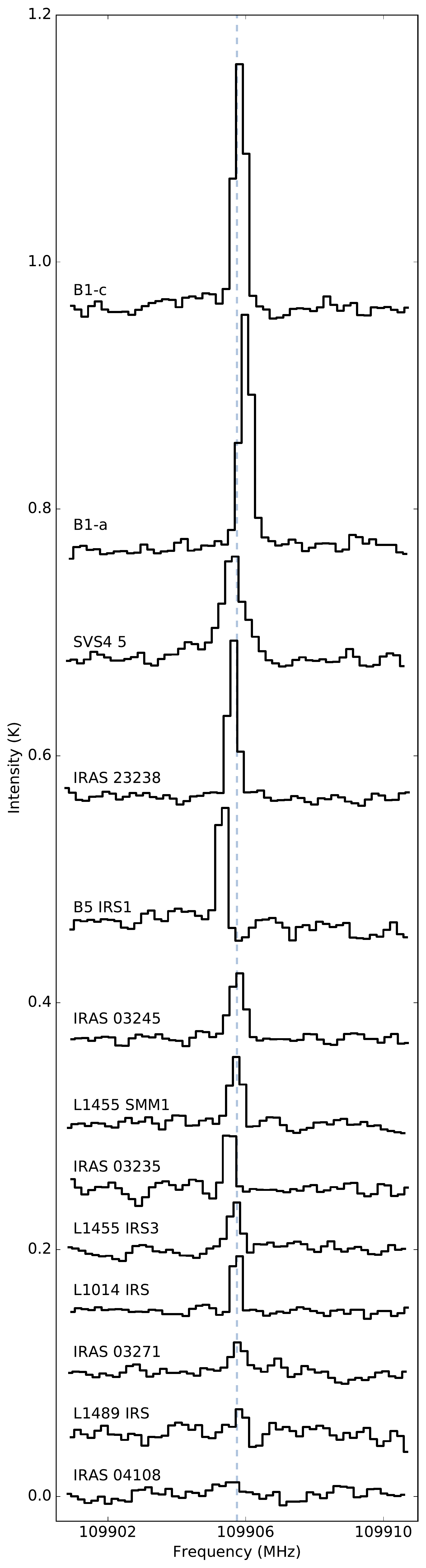}
	\caption{HNCO detected transitions.  Grey dashed lines show line centers; spectra are offset for clarity.}
	\label{spec_hnco}
\end{figure}

\begin{figure}[h!]
	\centering
	\includegraphics[width=0.8\linewidth]{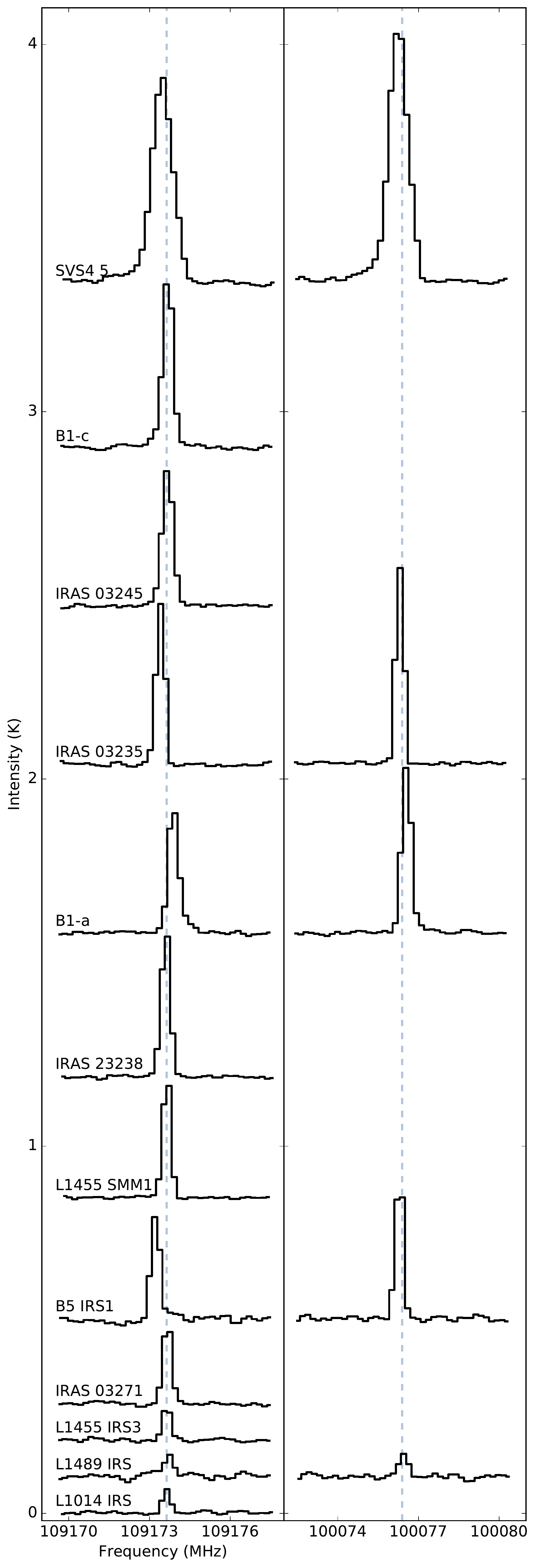}
	\caption{HC$_3$N detected transitions.  Grey dashed lines show line centers; spectra are offset for clarity.}
	\label{hc3n}
\end{figure}
\begin{figure}[h!]
	\centering
	\includegraphics[width=0.8\linewidth]{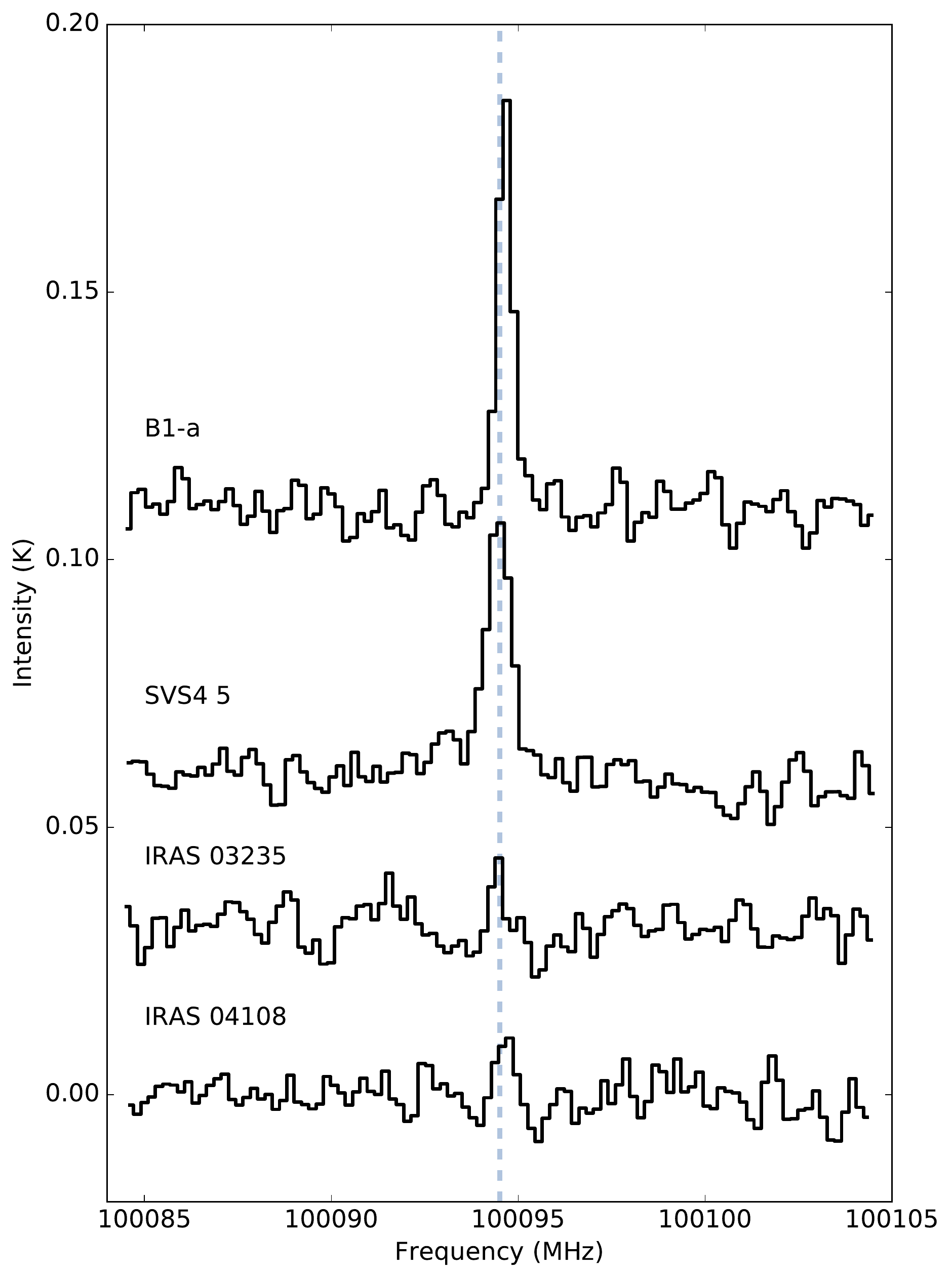}
	\caption{CH$_2$CO detected transitions.  Grey dashed lines show line centers; spectra are offset for clarity.}
	\label{spec_ch2co}
\end{figure}

\begin{figure*}[h!]
	\centering
	\includegraphics[width=0.8\textwidth]{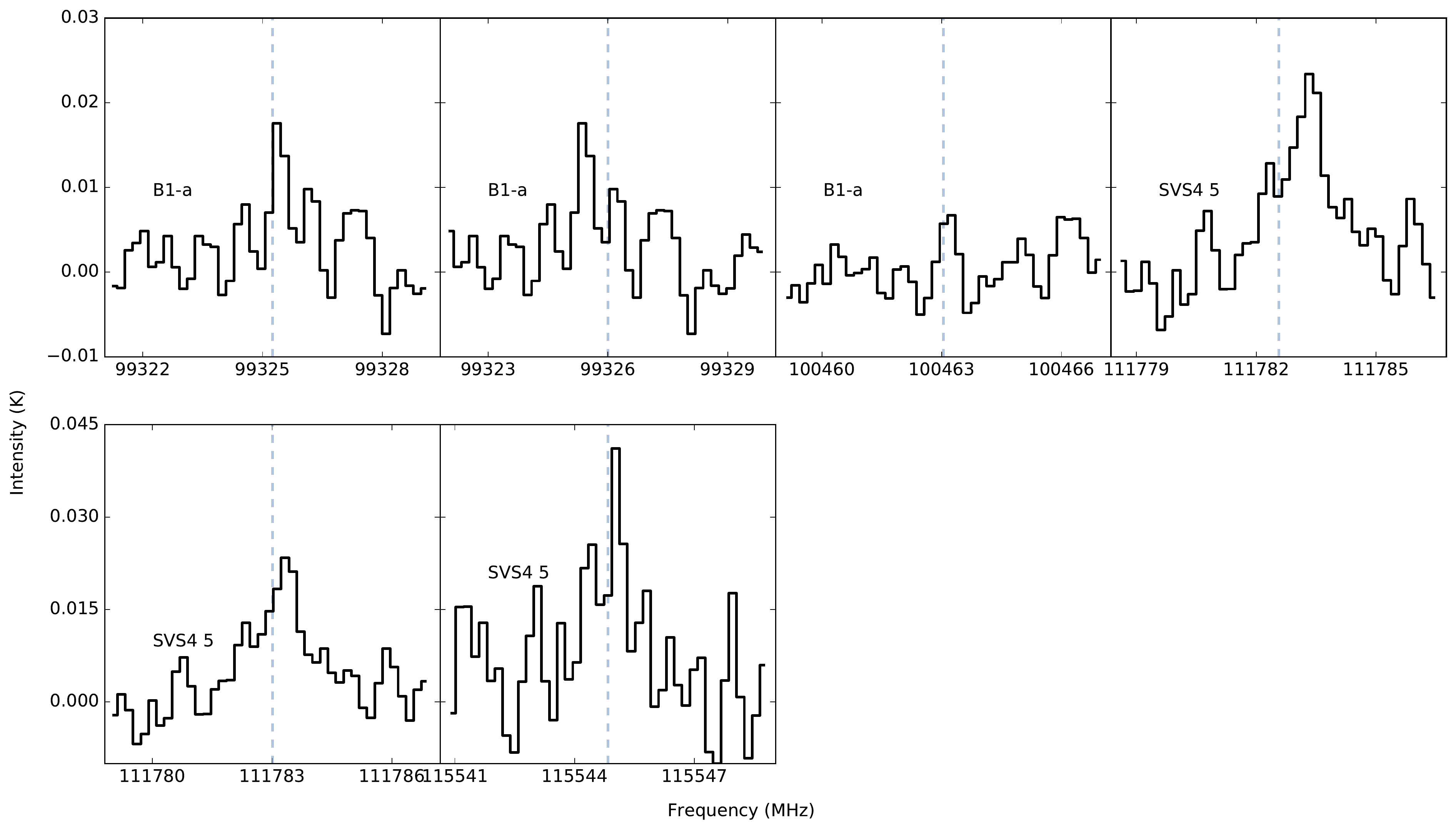}
	\caption{CH$_3$OCH$_3$ detected transitions.  Grey dashed lines show line centers; spectra are offset for clarity.}
	\label{spec_ch3och3}
\end{figure*}
\begin{figure*}[h!]
	\centering
	\includegraphics[width=0.8\textwidth]{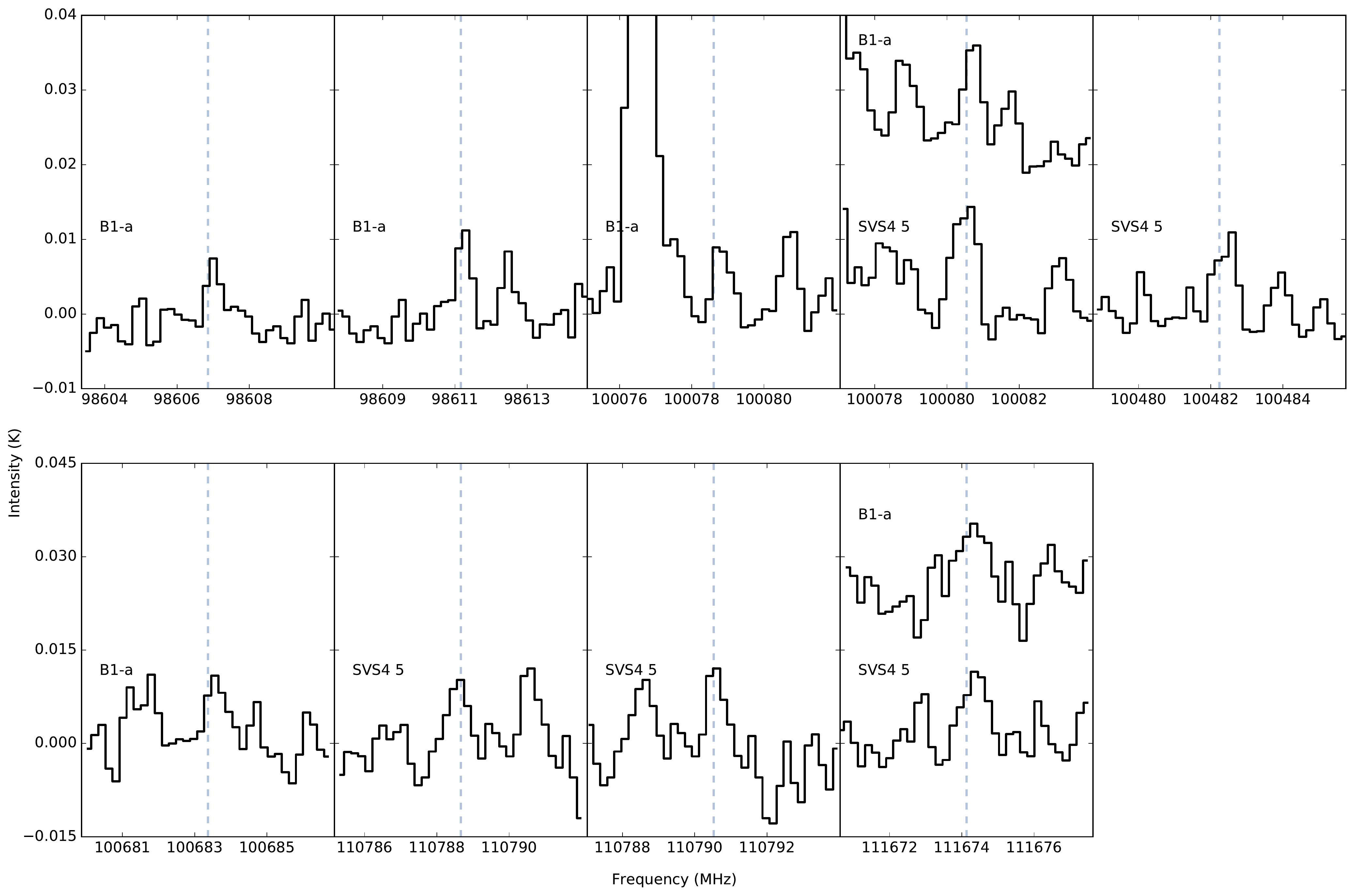}
	\caption{CH$_3$OCHO detected transitions.  Grey dashed lines show line centers; spectra are offset for clarity.}
	\label{spec_ch3ocho}
\end{figure*}

\clearpage

\section{Appendix: Rotation diagrams}
\begin{figure*}[]
	\centering
	\includegraphics[width=0.8\textwidth]{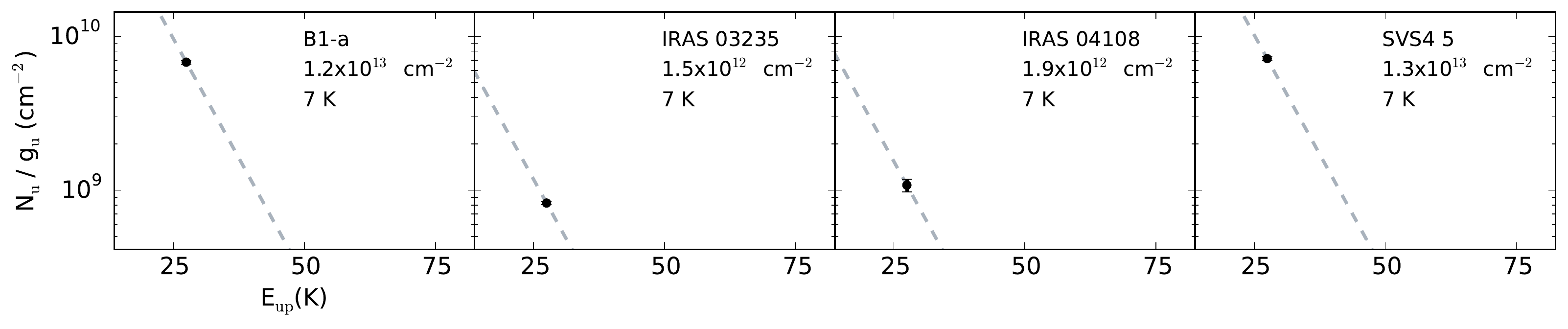}
	\caption{Rotation diagrams for CH$_{2}$CO.  Black circles indicate detections and grey triangles indicate upper limits.  Black dashed lines represent the fits to the data.  When a line could not be fit, a rotational temperature was assumed as described in the text, shown in grey dashed lines.}
	\label{RDCH2CO}
\end{figure*} 

\begin{figure*}[h!]
	\centering
	\includegraphics[width=0.8\textwidth]{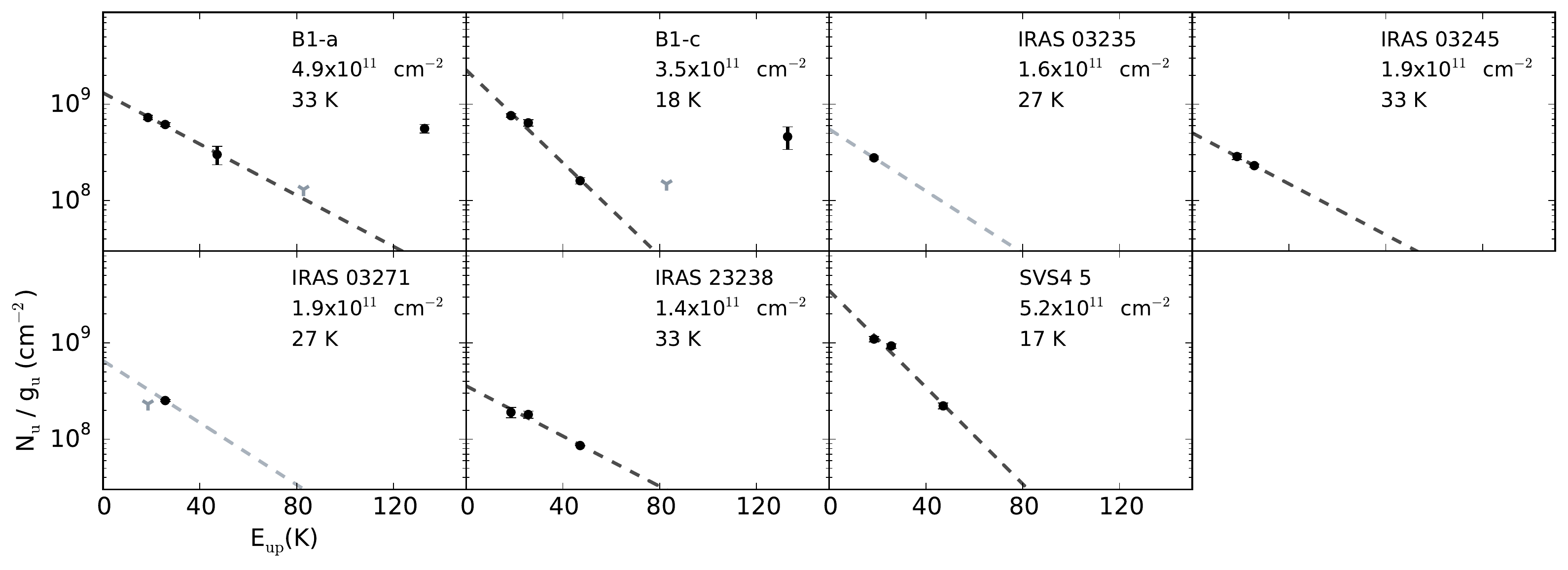}
	\caption{Rotation diagrams for CH$_{3}$CN.  Black circles indicate detections and grey triangles indicate upper limits.  Black dashed lines represent the fits to the data.  When a line could not be fit, a rotational temperature was assumed as described in the text, shown in grey dashed lines.}
	\label{RDCH3CN}
\end{figure*}

\begin{figure*}[h!]
	\centering
	\includegraphics[width=0.4\textwidth]{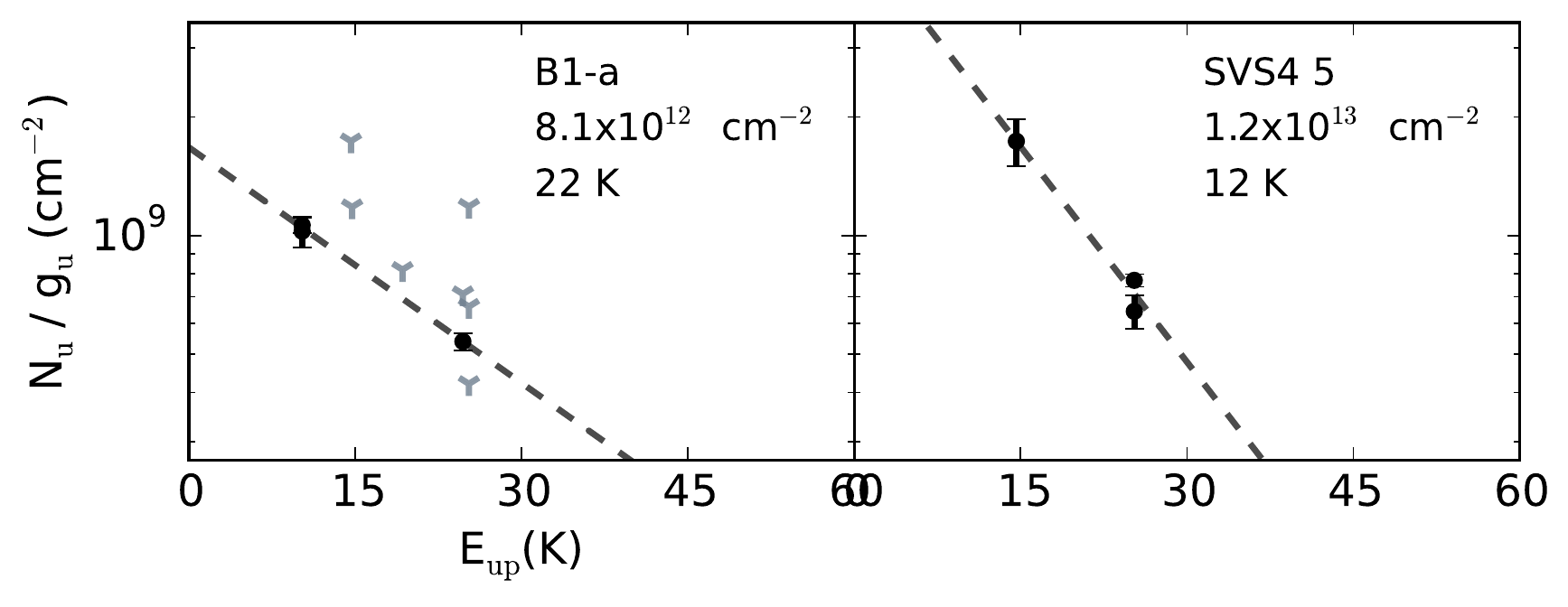}
	\caption{Rotation diagram for CH$_{3}$OCH$_{3}$.  Black circles indicate detections and grey triangles indicate upper limits.  Black dashed lines represent the fits to the data.  When a line could not be fit, a rotational temperature was assumed as described in the text, shown in grey dashed lines.}
	\label{RDCH3OCH3}
\end{figure*}
\begin{figure*}[h!]
	\centering
	\includegraphics[width=0.4\textwidth]{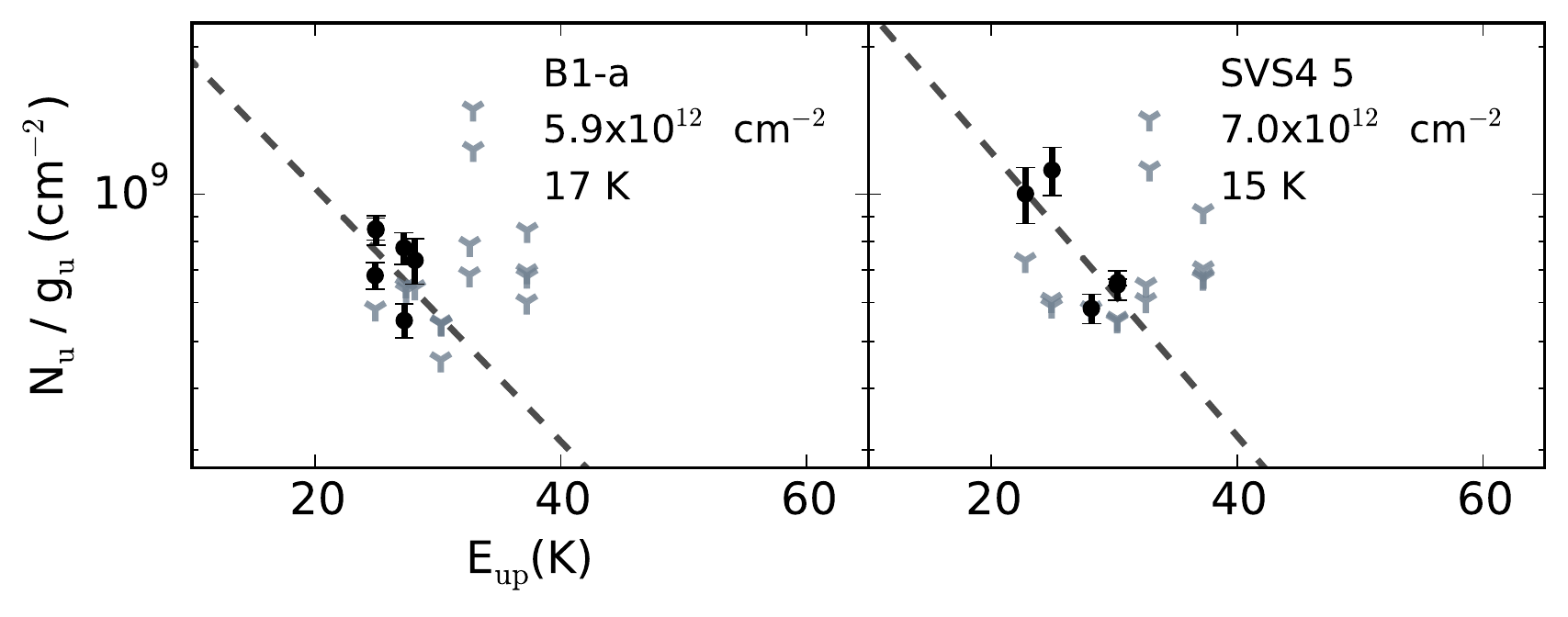}
	\caption{Rotation diagrams for CH$_{3}$OCHO.  Black circles indicate detections and grey triangles indicate upper limits.  Black dashed lines represent the fits to the data.  When a line could not be fit, a rotational temperature was assumed as described in the text, shown in grey dashed lines.}
	\label{RDCH3OCHO}
\end{figure*}
\begin{figure*}[h!]
	\centering
	\includegraphics[width=0.8\textwidth]{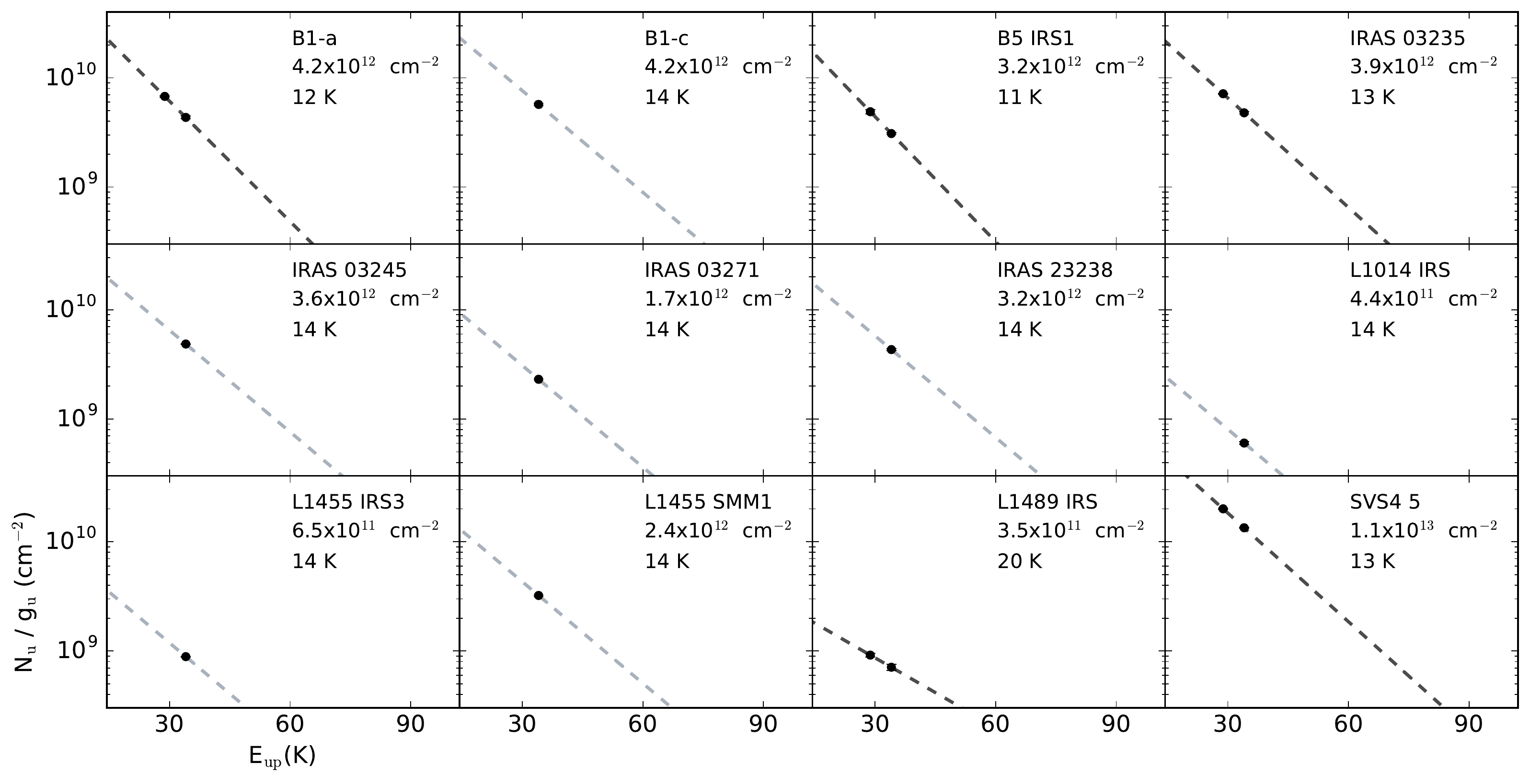}
	\caption{Rotation diagrams for HC$_{3}$N.  Black circles indicate detections and grey triangles indicate upper limits.  Black dashed lines represent the fits to the data.  When a line could not be fit, a rotational temperature was assumed as described in the text, shown in grey dashed lines.}
	\label{RDHC3n}
\end{figure*}
\begin{figure*}[h!]
	\centering
	\includegraphics[width=0.8\textwidth]{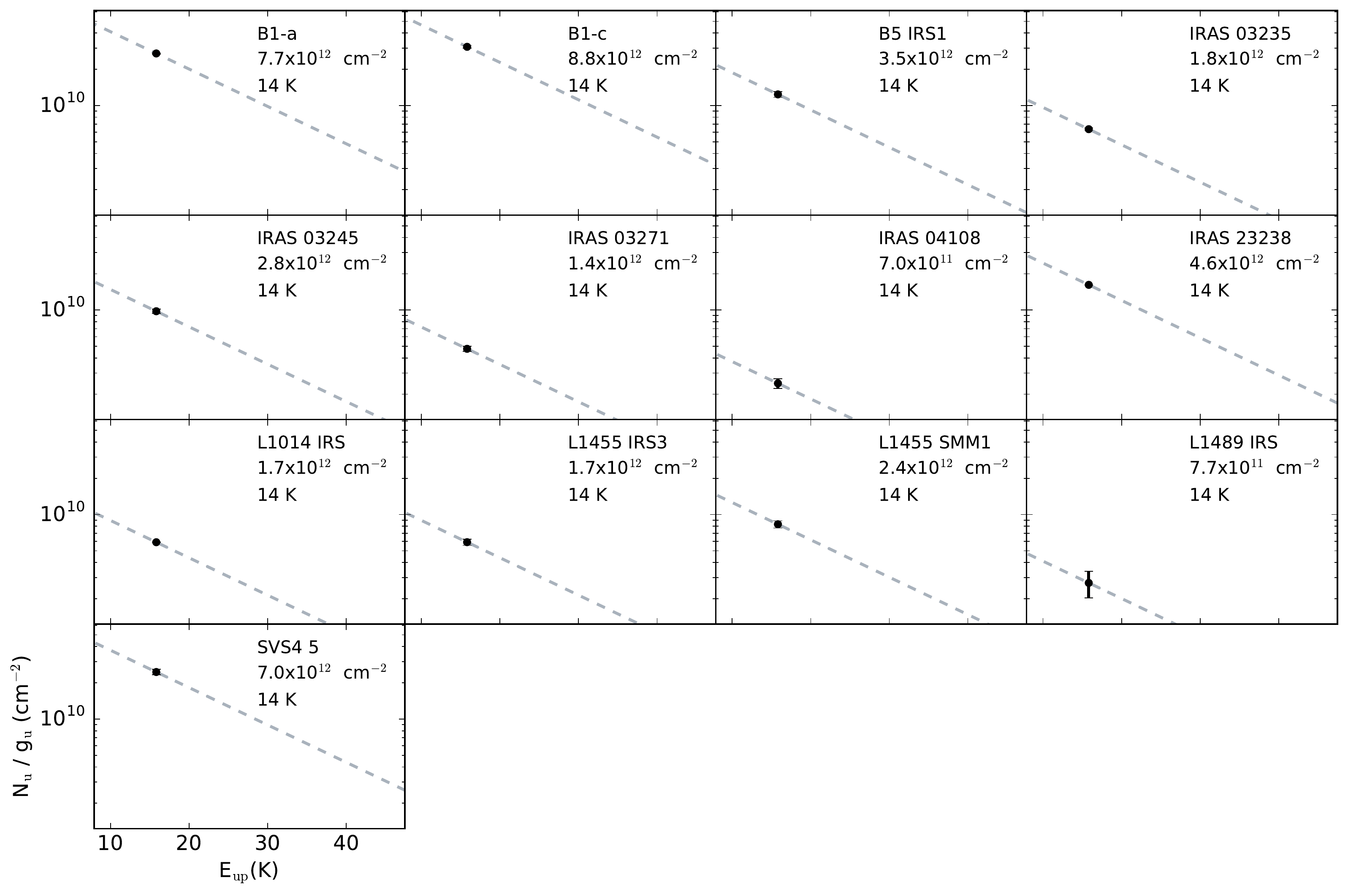}
	\caption{Rotation diagrams for HNCO.  Black circles indicate detections and grey triangles indicate upper limits.  Black dashed lines represent the fits to the data.  When a line could not be fit, a rotational temperature was assumed as described in the text, shown in grey dashed lines.}
	\label{RDHNCO}
\end{figure*}

\clearpage
\section{Appendix: Observed transitions and upper limits}
\begin{deluxetable*}{llllllll}
	\tabletypesize{\footnotesize}
	\tablecaption{Observed CH$_{2}$CO line intensities}
	\tablecolumns{3} 
	\tablehead{\colhead{Freq.}                             &
		\colhead{Transition}                               &  
		\multicolumn{6}{c}{$\int$T$_{\mathrm{MB}}$dV}     \\
		\colhead{(GHz)}                                    &
		\colhead{}                                         &
		\multicolumn{6}{c}{(K km s$^{-1}$)}                }
	\startdata
	&& B1-a & SVS 4-5 & B5 IRS1 & IRAS 03235 & IRAS 04108 & L1489 IRS  \\ \hline \hline
	100.095& 5$_{1,5}$ - 4$_{1,4}$ &0.117 (0.004)&0.124 (0.004)&$<$0.006&0.014 (0.000)&0.019 (0.002)&$<$0.006
	\enddata
	\label{ch2co_dets}
\end{deluxetable*}

\begin{deluxetable*}{clllllllll}
	\tabletypesize{\footnotesize}
	\tablecaption{Observed CH$_{3}$CHO line intensities}
	\tablecolumns{3} 
	\tablewidth{0.85\textwidth} 
	\tablehead{\colhead{Freq.}                                             &
		\colhead{Transition}                                               &
		\multicolumn{8}{c}{$\int$T$_{\mathrm{MB}}$dV }                    \\
		\colhead{(GHz)}                                                    &
		\colhead{}                                                         &
		\multicolumn{8}{c}{(K km s$^{-1}$)}                              }
	\startdata
	& & B1-a &  SVS 4-5 & B1-c & B5 IRS1 & HH 300 & IRAS 03235 & IRAS 03245 & IRAS 03254  \\ \hline \hline
	93.581& 5$_{1,5}$ - 4$_{1,4}$ A &0.104 (0.009)&0.107 (0.006)&-&-&-&-&-&-\\
	93.595& 5$_{1,5}$ - 4$_{1,4}$ E &0.099 (0.004)&0.124 (0.009)&0.023 (0.001)&-&-&-&-&-\\
	95.947& 5$_{0,5}$ - 4$_{0,4}$ E &0.159 (0.007)&0.190 (0.010)&-&-&-&-&0.018 (0.004)&-\\
	95.963& 5$_{0,5}$ - 4$_{0,4}$ A &0.172 (0.005)&0.197 (0.011)&0.033 (0.001)&$<$0.010&$<$0.004&$<$0.006&-&$<$0.006\\
	96.274& 5$_{2,4}$ - 4$_{2,3}$ A &0.036 (0.007)&$<$0.011&-&-&-&-&-&-\\
	96.426& 5$_{2,4}$ - 4$_{2,3}$ E &0.034 (0.006)&0.073 (0.007)&-&-&-&-&-&-\\
	96.476& 5$_{2,3}$ - 4$_{2,2}$ E &0.040 (0.005)&0.036 (0.006)&-&-&-&-&-&-\\
	96.633& 5$_{2,3}$ - 4$_{2,2}$ A &0.053 (0.011)&0.035 (0.000)&-&-&-&-&-&-\\
	98.863& 5$_{1,4}$ - 4$_{1,3}$ E &0.148 (0.004)&0.146 (0.007)&0.013 (0.003)&-&-&-&$<$0.004&-\\
	98.901& 5$_{1,4}$ - 4$_{1,3}$ A &0.144 (0.006)&0.159 (0.009)&$<$0.006&-&-&-&$<$0.003&-\\
	112.249& 6$_{1,6}$ - 5$_{1,5}$ A&0.098 (0.005)&0.108 (0.009)&$<$0.011&-&-&-&$<$0.007&-\\
	112.255& 6$_{1,6}$ - 5$_{1,5}$ E&0.110 (0.005)&0.104 (0.008)&$<$0.011&-&-&-&$<$0.008&-\\
	114.940& 6$_{0,6}$ - 5$_{0,5}$ E&0.135 (0.011)&0.204 (0.011)&$<$0.022&-&-&-&$<$0.009&-\\
	114.960& 6$_{0,6}$ - 5$_{0,5}$ A&0.135 (0.010)&0.143 (0.013)&$<$0.018&-&-&-&-&-\\
	115.494& 6$_{2,5}$ - 5$_{2,4}$ A&0.038 (0.009)&$<$0.023&-&-&-&-&-&-\\
	115.695& 6$_{2,5}$ - 5$_{2,4}$ E&0.071 (0.005)&$<$0.020&-&-&-&-&-&-\\
	115.910& 6$_{2,4}$ - 5$_{2,3}$ E&$<$0.018&$<$0.022&-&-&-&-&-&-\\
	116.118& 6$_{2,4}$ - 5$_{2,3}$ A&$<$0.021&0.059 (0.004)&-&-&-&-&-&-\\ \hline	
	& & IRAS 03271 & IRAS 04108 & IRAS 23238 & L1014 IRS & L1448 IRS1 & L1455 IRS3 & L1455 SMM1 & L1489 IRS     \\ \hline \hline
	93.581& 5$_{1,5}$ - 4$_{1,4}$ A  &-&-&-&-&-&-&-&-\\
	93.595& 5$_{1,5}$ - 4$_{1,4}$ E  &-&-&0.025 (0.000)&-&-&-&-&-\\
	95.947& 5$_{0,5}$ - 4$_{0,4}$ E  &-&-&0.035 (0.004)&-&-&-&-&-\\
	95.963& 5$_{0,5}$ - 4$_{0,4}$ A  &$<$0.004&$<$0.007&0.032 (0.002)&$<$0.003&$<$0.004&$<$0.004&-&$<$0.008\\
	98.863& 5$_{1,4}$ - 4$_{1,3}$ E   &-&-&0.021 (0.001)&-&-&-&-&-\\
	98.901& 5$_{1,4}$ - 4$_{1,3}$ A   &-&-&0.020 (0.000)&-&-&-&0.025 (0.004)&-\\
	112.249& 6$_{1,6}$ - 5$_{1,5}$ A  &-&-&$<$0.006&-&-&-&$<$0.008&-\\
	112.255& 6$_{1,6}$ - 5$_{1,5}$ E  &-&-&$<$0.005&-&-&-&0.022 (0.000)&-\\
	114.940& 6$_{0,6}$ - 5$_{0,5}$ E  &-&-&$<$0.008&-&-&-&$<$0.013&-\\
	114.960& 6$_{0,6}$ - 5$_{0,5}$ A  &-&-&$<$0.009&-&-&-&$<$0.013&-\\
	115.494& 6$_{2,5}$ - 5$_{2,4}$ A  &-&-&-&-&-&-&$<$0.015&-\\
	\enddata	
	\label{ch3chodets}
\end{deluxetable*}

\begin{deluxetable*}{clllllllllllllll}
	\tabletypesize{\footnotesize}
	\tablecaption{Observed CH$_{3}$CN line intensities}
	\tablecolumns{3} 
	\tablewidth{0.85\textwidth} 
	\tablehead{\colhead{Freq.}                                             &
		\colhead{Transition}                                               &
		\multicolumn{8}{c}{$\int$T$_{\mathrm{MB}}$dV }                    \\
		\colhead{(GHz)}                                                    &
		\colhead{}                                                         &
		\multicolumn{8}{c}{(K km s$^{-1}$)}                              }
	\startdata
	& & B1-a&  SVS 4-5 & B1-c & B5 IRS1 & HH 300 & IRAS 03235 & IRAS 03245 & IRAS 03254  \\ \hline \hline
	110.349& 6$_{4}$ - 5$_{4}$ 110.349& 6(4) - 5(4) &0.026 (0.003)&-&0.022 (0.006)&-&-&-&-&-\\
	110.364& 6$_{3}$ - 5$_{3}$ 110.364& 6(2) - 5(2) &$<$0.005&-&$<$0.006&-&-&-&-&-\\
	110.375& 6$_{2}$ - 5$_{2}$ 110.375& 6(1) - 5(1) &0.022 (0.005)&0.017 (0.001)&0.012 (0.001)&-&-&-&-&-\\
	110.381& 6$_{1}$ - 5$_{1}$ 110.381& 6(0) - 5(0) &0.050 (0.002)&0.076 (0.004)&0.052 (0.004)&-&-&-&0.019 (0.001)&-\\
	110.383& 6$_{0}$ - 5$_{0}$ 110.383& 6(4) - 5(4) &0.061 (0.003)&0.092 (0.006)&0.064 (0.003)&$<$0.008&$<$0.006&0.023 (0.001)&0.024 (0.002)&$<$0.005\\ \hline
	& & IRAS 03271 & IRAS 04108 & IRAS 23238 & L1014 IRS & L1448 IRS1 & L1455 IRS3 & L1455 SMM1 & L1489 IRS     \\ \hline \hline
	110.375& 6$_{2}$ - 5$_{2}$ &-&-&0.006 (0.000)&-&-&-&-&-\\
	110.381& 6$_{1}$ - 5$_{1}$ &0.021 (0.000)&-&0.015 (0.001)&-&-&-&-&-\\
	110.383& 6$_{0}$ - 5$_{0}$ &$<$0.007&$<$0.006&0.016 (0.002)&$<$0.003&$<$0.005&$<$0.004&$<$0.005&$<$0.007\\
	\enddata	
	\label{ch3cndets}	
\end{deluxetable*}

\begin{deluxetable*}{clllllllllllllll}
	\tabletypesize{\footnotesize}
	\tablecaption{Observed CH$_{3}$OCH$_3$ line intensities}
	\tablecolumns{3} 
	\tablewidth{0.85\textwidth} 
	\tablehead{\colhead{Freq.}                                             &
		\colhead{Transition}                                               &
		\multicolumn{8}{c}{$\int$T$_{\mathrm{MB}}$dV }                    \\
		\colhead{(GHz)}                                                    &
		\colhead{}                                                         &
		\multicolumn{8}{c}{(K km s$^{-1}$)}                              }
	\startdata
& & B1-a &  SVS 4-5 & B1-c & B5 IRS1 & HH 300 & IRAS 03235 & IRAS 03245 & IRAS 03254  \\ \hline \hline
93.857& 4$_{2,3}$ - 4$_{1,4}$ EE &$<$0.005&-&-&-&-&-&-&-\\
96.850& 5$_{2,4}$ - 5$_{1,5}$ EE &$<$0.004&-&-&-&-&-&-&-\\
99.325& 4$_{1,4}$ - 3$_{0,3}$ EE &0.017 (0.001)&-&-&-&-&-&-&-\\
99.326& 4$_{1,4}$ - 3$_{0,3}$ AA &0.011 (0.001)&-&$<$0.006&$<$0.006&$<$0.004&$<$0.004&$<$0.003&$<$0.005\\
100.463& 6$_{2,5}$ - 6$_{1,6}$ EE&0.010 (0.000)&-&-&-&-&-&-&-\\
100.466& 6$_{2,5}$ - 6$_{1,6}$ AA&$<$0.003&-&-&-&-&-&-&-\\
111.783& 7$_{0,7}$ - 6$_{1,6}$ AA&$<$0.004&0.012 (0.001)&-&-&-&-&-&-\\
111.783& 7$_{0,7}$ - 6$_{1,6}$ EE&$<$0.004&0.022 (0.001)&-&-&-&-&-&-\\
111.784& 7$_{0,7}$ - 7$_{1,6}$ AE&$<$0.004&-&-&-&-&-&-&-\\
115.545& 5$_{1,5}$ - 4$_{0,4}$ EE&$<$0.014&0.041 (0.006)&-&-&-&-&-&-\\
\hline
& & IRAS 03271 & IRAS 04108 & IRAS 23238 & L1014 IRS & L1448 IRS1 & L1455 IRS3 & L1455 SMM1 & L1489 IRS   \\ \hline \hline
99.326& 4$_{1,4}$ - 3$_{0,3}$ AA &$<$0.005&$<$0.004&$<$0.003&$<$0.002&$<$0.003&$<$0.005&$<$0.004&$<$0.005\\
	\enddata
	\label{ch3och3dets}
\end{deluxetable*}

\begin{deluxetable*}{clllllllllllllll}
	\tabletypesize{\footnotesize}
	\tablecaption{Observed CH$_{3}$OCHO line intensities}
	\tablecolumns{3} 
	\tablewidth{0.85\textwidth} 
	\tablehead{\colhead{Freq.}                                             &
		\colhead{Transition}                                               &
		\multicolumn{8}{c}{$\int$T$_{\mathrm{MB}}$dV }                    \\
		\colhead{(GHz)}                                                    &
		\colhead{}                                                         &
		\multicolumn{8}{c}{(K km s$^{-1}$)}                              }
	\startdata
	& & B1-a & SVS 4-5 & B1-c & B5 IRS1 & HH 300 & IRAS 03235 & IRAS 03245 & IRAS 03254 \\ \hline \hline
	98.607&  8$_{3,6}$ - 7$_{3,5}$   E &0.012 (0.001)&-&-&-&-&-&-&-\\
	98.611&  8$_{3,6}$ - 7$_{3,5}$   A &0.017 (0.001)&-&$<$0.008&$<$0.007&$<$0.005&$<$0.003&$<$0.005&$<$0.004\\
	100.079&  9$_{1,9}$ - 8$_{1,8}$  E &0.024 (0.002)&$<$0.007&-&-&-&-&-&-\\
	100.081&  9$_{1,9}$ - 8$_{1,8}$  A &0.024 (0.001)&0.032 (0.004)&-&-&-&-&-&-\\
	100.295&  8$_{3,5}$ - 7$_{3,4}$  E &$<$0.005&-&-&-&-&-&-&-\\
	100.308&  8$_{3,5}$ - 7$_{3,4}$  A &$<$0.005&-&-&-&-&-&-&-\\
	100.482&  8$_{1,7}$ - 7$_{1,6}$  E &-&0.025 (0.003)&-&-&-&-&-&-\\
	100.491&  8$_{1,7}$ - 7$_{1,6}$  A &-&$<$0.006&-&-&-&-&-&-\\
	100.682&  9$_{0,9}$ - 8$_{0,8}$  E &$<$0.006&$<$0.006&-&-&-&-&-&-\\
	100.683&  9$_{0,9}$ - 8$_{0,8}$  A &0.019 (0.001)&$<$0.006&-&-&-&-&-&-\\
	110.789& 10$_{1,10}$ - 9$_{1,9}$ E &$<$0.006&0.023 (0.000)&-&-&-&-&-&-\\
	110.791& 10$_{1,10}$ - 9$_{1,9}$ A &$<$0.006&0.023 (0.002)&-&-&-&-&-&-\\
	110.880&  9$_{5,5}$ - 8$_{5,4}$  A &$<$0.007&$<$0.006&-&-&-&-&-&-\\
	110.887&  9$_{3,7}$ - 8$_{3,6}$  A &$<$0.006&$<$0.006&-&-&-&-&-&-\\
	111.170& 10$_{0,10}$ - 9$_{0,9}$ E &$<$0.006&$<$0.006&-&-&-&-&-&-\\
	111.172& 10$_{0,10}$ - 9$_{0,9}$ A &$<$0.005&$<$0.006&-&-&-&-&-&-\\
	111.196&  9$_{4,6}$ - 8$_{4,5}$  A &$<$0.005&$<$0.006&-&-&-&-&-&-\\
	111.223&  9$_{4,6}$ - 8$_{4,5}$  E &$<$0.006&$<$0.007&-&-&-&-&-&-\\
	111.408&  9$_{4,5}$ - 8$_{4,4}$  E &$<$0.007&$<$0.005&-&-&-&-&-&-\\
	111.453&  9$_{4,5}$ - 8$_{4,4}$  A &$<$0.006&$<$0.006&-&-&-&-&-&-\\
	111.674&  9$_{1,8}$ - 8$_{1,7}$  E &0.023 (0.002)&0.018 (0.001)&-&-&-&-&-&-\\
	111.682&  9$_{1,8}$ - 8$_{1,7}$  A &$<$0.007&$<$0.006&-&-&-&-&-&-\\
	113.743&  9$_{3,6}$ - 8$_{3,5}$  E &$<$0.012&$<$0.011&-&-&-&-&-&-\\
	113.757&  9$_{3,6}$ - 8$_{3,5}$  A &$<$0.014&$<$0.014&-&-&-&-&-&-\\
	116.545&  9$_{2,7}$ - 8$_{2,6}$  E &$<$0.029&$<$0.025&-&-&-&-&-&-\\
	116.558&  9$_{2,7}$ - 8$_{2,6}$  A &$<$0.033&$<$0.027&-&-&-&-&-&-\\ \hline
	& &  IRAS 03271 & IRAS 04108 & IRAS 23238 & L1014 IRS & L1448 IRS1 & L1455 IRS3 & L1455 SMM1 & L1489 IRS  \\ \hline \hline
	98.611&8$_{3,6}$ - 7$_{3,5}$ A &$<$0.004&$<$0.005&$<$0.003&$<$0.003&$<$0.003&$<$0.004&$<$0.004&$<$0.005\\	\enddata
	\label{ch3ochodets}
\end{deluxetable*}

\begin{deluxetable*}{clllllllllllllll}
	\tabletypesize{\footnotesize}
	\tablecaption{Observed HC$_{3}$N line intensities}
	\tablecolumns{3} 
	\tablewidth{0.85\textwidth} 
	\tablehead{\colhead{Freq.}                                             &
		\colhead{Transition}                                               &
		\multicolumn{8}{c}{$\int$T$_{\mathrm{MB}}$dV }                    \\
		\colhead{(GHz)}                                                    &
		\colhead{}                                                         &
		\multicolumn{8}{c}{(K km s$^{-1}$)}                              }
	\startdata
	& & B1-a & SVS 4-5 & B1-c & B5 IRS1 & HH 300 & IRAS 03235 & IRAS 03245 & IRAS 03254 \\ \hline \hline
	100.076& 11-10 &0.620 (0.010)&1.839 (0.029)&-&0.449 (0.020)&-&0.657 (0.007)&-&-\\
	109.174& 12-11 &0.474 (0.014)&1.467 (0.019)&0.624 (0.014)&0.337 (0.006)&$<$0.007&0.522 (0.015)&0.531 (0.004)&$<$0.005\\\hline
	& &  IRAS 03271 & IRAS 04108 & IRAS 23238 & L1014 IRS & L1448 IRS1 & L1455 IRS3 & L1455 SMM1 & L1489 IRS  \\ \hline \hline	
	100.076&11-10 &-&-&-&-&-&-&-&0.084 (0.003)\\
	109.174&12-11 &0.253 (0.004)&$<$0.005&0.472 (0.009)&0.066 (0.002)&$<$0.004&0.097 (0.001)&0.352 (0.007)&0.078 (0.005)\\
	\enddata
	\label{hc3ndets}
	
\end{deluxetable*}

\begin{deluxetable*}{clllllllllllllll}
	\tabletypesize{\footnotesize}
	\tablecaption{Observed HNCO line intensities}
	\tablecolumns{3} 
	\tablewidth{0.85\textwidth} 
	\tablehead{\colhead{Freq.}                                             &
		\colhead{Transition}                                               &
		\multicolumn{8}{c}{$\int$T$_{\mathrm{MB}}$dV }                    \\
		\colhead{(GHz)}                                                    &
		\colhead{}                                                         &
		\multicolumn{8}{c}{(K km s$^{-1}$)}                              }
	\startdata
	& & B1-a & SVS 4-5 & B1-c & B5 IRS1 & HH 300 & IRAS 03235 & IRAS 03245 & IRAS 03254 \\ \hline \hline
	109.906& 5$_{0,5}$ - 4$_{0,4}$ &0.228 (0.002)&0.191 (0.010)&0.247 (0.007)&0.097 (0.005)&$<$0.006&0.050 (0.001)&0.076 (0.003)&$<$0.006\\ \hline
	& & IRAS 03271 & IRAS 04108 & IRAS 23238 & L1014 IRS & L1448 IRS1 & L1455 IRS3 & L1455 SMM1 & L1489 IRS   \\ \hline \hline
	109.906& 5$_{0,5}$ - 4$_{0,4}$ &0.037 (0.002)&0.019 (0.002)&0.126 (0.000)&0.046 (0.000)&$<$0.004&0.046 (0.003)&0.064 (0.004)&0.021 (0.005)\\
	\enddata
	\label{hnco_dets}
\end{deluxetable*}

\bibliographystyle{apj}

\end{document}